\documentclass[10pt]{iopart}
\usepackage{amssymb,amsthm}
\usepackage{graphicx}% Include figure files
\usepackage{dcolumn}% Align table columns on decimal point
\usepackage{bm}% bold math
\usepackage{bbm}
\usepackage{enumitem}
\usepackage{cite}
\usepackage{tikz}
\usepackage{color}
\usepackage{hyperref}
\usepackage{footmisc}
\usepackage[normalem]{ulem}

\definecolor{myrefcolor}{rgb}{0,0.5,0}
\hypersetup{
    colorlinks=true,
    citecolor=red,
    filecolor=black,
    linkcolor=myrefcolor,
    urlcolor=black
}
\newcommand{\beq}[0]{\begin{equation}}
\newcommand{\eeq}[0]{\end{equation}}
\newcommand{\bw}[0]{\begin{widetext}}
\newcommand{\ew}[0]{\end{widetext}}
\newcommand{\bc}[0]{\begin{center}}
\newcommand{\ec}[0]{\end{center}}
\newcommand{\bwn}[0]{\begin{widetext}\begin{eqnarray}}
\newcommand{\ewn}[0]{\end{eqnarray}\end{widetext}}
\newcommand{\beqn}[0]{\begin{eqnarray}}
\newcommand{\eeqn}[0]{\end{eqnarray}}

\newcommand{\proj}[1]{|#1\rangle \!\langle #1|}
\newcommand{\ket}[1]{|#1\rangle}
\newcommand{\bra}[1]{\langle #1 |}

\newcommand{\cee}[1]{\mathbb{C}^{#1}}

\newcommand{\jedynka}[0]{\mathbbm{1}}

\newcommand{\non}[0]{\nonumber\\}

\newcommand{\linka}[0]{\noindent\makebox[\linewidth]{\rule{\textwidth}{0.4pt}}}

\newcommand{\grod}[0]{Grothendieck\;}
\newcommand{\numerek}[1]{$\langle #1 \rangle$}

\def\etal{{\it et al.\;}}
\def\calA{{\cal A}}
\def\calB{{\cal B}}
\def\calC{{\cal C}}

\def\calI{{\cal I}}

\def\calM{{\cal M}}
\def\calN{{\cal N}}

\def\calP{{\cal P}}
\def\calQ{{\cal Q}}

\newcommand{\acin}[0]{Ac\'{i}n\;}

\newcommand{\be}{\begin{eqnarray}}
\newcommand{\ee}{\end{eqnarray}}

\newcommand{\ba}{\begin{array}}
\newcommand{\ea}{\end{array}}

\newcommand{\ot}[0]{\otimes}

\theoremstyle{plain}
\newtheorem{thm}{Theorem}%[section]

\theoremstyle{definition}

\theoremstyle{remark}
\newtheorem{rem}[thm]{Remark}
\begin{document}

\title[Local hidden--variable models ...]{Local hidden--variable models for
entangled quantum states}

\author{R Augusiak$^1$, M Demianowicz$^1$, and A Ac\'{i}n$^{1,2}$}
\address{$^1$ ICFO--Institut de Ciencies Fotoniques, Mediterranean Technology
Park, 08860 Castelldefels (Barcelona), Spain}
\address{$^2$ ICREA--Institucio Catalana de Recerca i
Estudis Avan\c{c}ats, Lluis Companys 23, 08010 Barcelona, Spain}

\begin{abstract}
While entanglement and violation of Bell inequalities were
initially thought to be equivalent quantum phenomena, we now have
different examples of entangled states whose correlations can be
described by local hidden--variable models and, therefore, do not
violate any Bell inequality. We provide an up to date overview of
the existing local hidden--variable models for entangled quantum
states, both in the bipartite and multipartite case, and discuss
some of the most relevant open questions in this context. Our
review covers twenty five years of this line of research since the
seminal work by Werner [R. F. Werner, Phys. Rev. A \textbf{40}, 8
(1989)] providing the first example of an entangled state with a
local model, which in turn appeared twenty five years after the
seminal work by Bell [J. S. Bell, Physics \textbf{1}, 195 (1964)],
about the impossibility of recovering the predictions of quantum
mechanics using a local hidden--variables theory.
\end{abstract}

%Uncomment for PACS numbers title message
%\pacs{00.00, 20.00, 42.10}
% Keywords required only for MST, PB, PMB, PM, JOA, JOB?
%\vspace{2pc}
%\noindent{\it Keywords}: Article preparation, IOP journals
% Uncomment for Submitted to journal title message
%\submitto{\JPA}
% Comment out if separate title page not required
\maketitle \tableofcontents
\section{Introduction}

In 1935, the concepts of local--hidden variable model and
entanglement were introduced in the works by, respectively,
Einstein, Podolsky and Rosen \cite{epr}, and Schr\"odinger
\cite{schrodinger}. After them, very few works (see, e.g.,
\cite{BohmI,BohmII}) considered the problem of whether local
models may provide a more intuitive, and complete, alternative to
the quantum formalism until 1964, when Bell showed that local
models are in fact in contradiction with quantum predictions. In
particular, he proved that local models satisfy some inequalities,
known thereafter as Bell inequalities, that are violated by the
statistics of local measurements on a singlet state \cite{Bell}.

The work of Bell started the study of quantum nonlocality, that
is, of those correlations obtained when performing local
measurements on entangled states that do not have a classical
analogue. Initially, it was believed that entanglement and quantum nonlocality
were equivalent phenomena. And, in fact, entanglement and
nonlocality do coincide for pure states, as any pure entangled
state violates a Bell inequality \cite{gisin}. However, in 1989 Werner
introduced a family of highly symmetric mixed entangled states, known
today as the Werner states, and exploited the symmetries of these states to
construct an explicit local model reproducing the correlations for some of them
\cite{Werner}. The work by Werner then implied that the relation between
nonlocality and entanglement is subtler than expected and, in particular, these
two notions do not coincide in the standard scenario originally introduced by
Bell.

After Werner's work, several other results have appeared providing new local
models for other
entangled states. The main purpose of this article is to review all these
existing models. Our hope is that this review will be useful to have a broad
vision of what is known today on the relation between entanglement and quantum
nonlocality, in particular in the case when an entangled
state satisfies all Bell inequalities. As it will become clear below, the
explicit construction of local models has turned out to be an extremely
difficult problem and, at the moment, we only have a few models beyond Werner's
original construction. In fact, in our view, most, if not all of the existing
models, can be interpreted, in a way or another, as variants of Werner's model.

 The structure of the review is the following: after introducing the main concepts and definitions
 used in our work, we go through all the existing models, both in the bipartite and multipartite scenario.
  We then briefly discuss other possible definitions of nonlocality, such as, for instance, the concept of
  hidden nonlocality introduced by Popescu \cite{popescu}. Finally, we present our conclusions and a list of
   open questions.

\section{Preliminaries and notation}\label{preliminaries}

We start by introducing notions and definitions repeatedly used
throughout the manuscript: entanglement and genuine multipartite
entanglement, measurement in quantum theory, and the
Bell--experiment setup with the corresponding notions of locality.

In what follows by $B(\mathcal{H})$, $\mathbbm{1}_d$, $\Omega_d$, and $\omega_d$
we denote, respectively, the set of bounded linear operators acting on a
finite-dimensional Hilbert space, the $d\times d$ identity
matrix, the set
$\Omega_d=\{|\lambda\rangle\in\mathbbm{C}^d|\langle\lambda|\lambda\rangle=1\} $,
and the unique distribution over $\Omega_d$ invariant under any unitary
operation $U$ acting on $\mathbbm{C}^d$.

\subsection{Entanglement}\label{splatanie}

Let us consider two parties, traditionally called Alice ($A$) and
Bob ($B$), sharing a bipartite quantum state $\rho_{AB}$ acting on
a product Hilbert space $\mathcal{H}=\mathbbm{C}^{d_A}\ot\mathbbm{C}^{d_B}$.  We
say that $\rho_{AB}$ is {\it separable} iff it can be written as a convex
combination of pure product states \cite{Werner}, that is
\beq\label{separowalny} \rho_{AB}=\sum_i p_i
\proj{\psi_A^i}\otimes \proj{\phi_B^i},\quad p_i\geq0,\quad\sum_i p_i =1 \eeq
with some $\ket{\psi_A^i}\in \cee{d_A}$ and $\ket{\phi_B^i}\in
\cee{d_B}$. Otherwise, it is called
{\it entangled}. In a general multiparty scenario with $N$ parties
we will denote them $A^{(1)},\ldots,A^{(N)}=:\boldsymbol{A}$. Let
$\boldsymbol{A}_k$ be a $k$ element subset of
$\boldsymbol{A}$ and $\overline{\boldsymbol{A}}_k$ its complement
in the said set. Then, an $N$ partite state
$\rho_{\boldsymbol{A}}$ is said to be {\it biseparable} if it can
be written as
\beq\label{biseparowalny}
\rho_{\boldsymbol{A}}=\sum_{\boldsymbol{A}_k|\overline{\boldsymbol{A}}_k}
p_{\boldsymbol{A}_k|\overline{\boldsymbol{A}}_k}
\rho_{\boldsymbol{A}_k|\overline{\boldsymbol{A}}_k}, \quad
\sum_{\boldsymbol{A}_k|\overline{\boldsymbol{A}}_k}
p_{\boldsymbol{A}_k|\overline{\boldsymbol{A}}_k}=1, \eeq
where each $\rho_{\boldsymbol{A}_k|\overline{\boldsymbol{A}}_k}$
is separable (see (\ref{separowalny})) across the {\it
bipartition} $\boldsymbol{A}_k|\overline{\boldsymbol{A}}_k$ and
the sum goes over all such bipartitions. If a state cannot be
written as (\ref{biseparowalny}) it is called {\it genuinely
multiparty entangled} (GME).

\subsection{Measurement in quantum theory}
\label{Sec:Measurement}

We call a collection $\{P_a\}_{a=0}^{k-1}$ of $k$ projections
acting on $\mathbbm{C}^d$ a \textit{$k$-outcome projective
measurement} (PM; also called \textit{von Neumann measurement}) if
the $P_a$ are supported on orthogonal subspaces, i.e.,
$P_aP_{a'}=P_a\delta_{aa'}$, and
\begin{equation}\label{norm}
\sum_{a=0}^{k-1}P_a=\mathbbm{1}_d,
\end{equation}
where by $\mathbbm{1}_d$ we denote the $d\times d$ identity
operator. With a projective measurement with outcomes $\alpha_a$,
$a=0,1,\cdots,k-1$, we associate an operator
$\calA=\sum_{a=0}^{k-1} \alpha_a P_a$ called an
\textit{observable} which is measured in the measurement process.
In what follows, we just talk about performing measurement
$\calA$. Performed on a state $\varrho$, a measurement will give
the $a$th outcome $\alpha_a$ corresponding to $P_a$ with the
probability $p(a|\calA)$ given by the Born rule
\beq\label{born} p(a|\calA)= \tr (P_a \varrho).\eeq
The mean value of an observable in the state $\varrho$ is given by
\beq\langle \calA \rangle :=\tr (\calA\varrho)= \sum_{a=0}^{k-1}
\alpha_a\, p(a|\calA). \eeq

Clearly, when $d=2$ a projective measurement can have only two
outcomes represented by rank-one projections $P_a$ (of course,
there is also a trivial single-outcome projective measurement with
$P_0=\mathbbm{1}_d$, which we do not need to consider here). In
general, however, the projectors $P_a$ do not necessarily have to
be rank one, and, in fact, if $k<d$ there must exist at least one
of rank larger than one.

A collection of $k$ operators $\calA=\{\calA_a\}_{a=0}^{k-1}$
acting on $\mathbbm{C}^d$ is called a \textit{$k$-outcome POVM},
where POVM stands for Positive-Operator Valued Measure, or
\textit{$k$-outcome generalized measurement} if they are positive,
i.e., $\calA_a\geq 0$, and
\begin{equation}\label{norm2}
\sum_{a=0}^{k-1} \calA_a=\mathbbm{1}_d.
\end{equation}
The latter, in particular, means that every $\mathcal{A}_a$ is
upper bounded by the identity, i.e.,
$\mathcal{A}_a\leq\mathbbm{1}_d$. The probability of obtaining the
outcome $a$ corresponding to $\calA_a$ is given again by the Born
rule, that is, by Eq. (\ref{born}) with $P_a$ replaced by
$\calA_a$.

 In what follows the operators $P_a$ forming a
projective measurement as well as $\calA_a$ forming a generalized
measurement will be called \textit{measurement operators}.
Obviously, every projective measurement is also a POVM, the
opposite, however, is in general not true. Equivalently speaking,
for a given dimension $d$, the set of all projective measurements
forms a proper subset of all generalized measurements
\footnote{The restriction to the fixed dimensions is very
important here since any POVM can be realized formally as a PM in
a larger space.}. It is important to note, however, that whenever
we deal with $d$ outcome measurements on $\cee{d}$ they are
necessarily projective.
For further benefits, we also notice that in our considerations we
can in fact assume that  measurement operators, both in the case
of projective and generalized measurements, are rank-one, or, in
other words, the measurements are nondegenerate. This is because
any measurement whose operators are of rank larger than one can
always be realized as a measurement with rank-one operators. More
precisely, since every element $\calA_a$ of a generalized
measurement is a positive operator it admits the form
$\calA_a=\sum_i {\eta}_i^{(a)}P_i^{(a)}$ with $0\leq
\eta_i^{(a)}\leq 1$ and $P_i^{(a)}$ being, respectively, the
eigenvalues and the rank-one eigenprojectors of $\calA_a$.
Moreover, due to Eq. (\ref{norm2}), all the eigenvalues must
satisfy
\beq\label{condition} \sum_{i,a}\eta_i^{(a)}=d. \eeq
Now, denoting all these new rank-one
operators as $\calA_{a,i}={\eta}_i^{(a)}P_i^{(a)}$ it is clear
that they also form a POVM which is a ``finer grained" version of
$\{\calA_a\}$. To reproduce the statistics of the ``coarse
grained'' original POVM, one simply applies the finer POVM
$\{\calA_{a,i}\}$ and forgets the result $i$. In this way, a
local model (see the upcoming section for the relevant notion) for
POVMs with rank-one measurement operators will imply a local model
for any POVM.

It is also worth mentioning that whenever we work with qubits it
is beneficial to exploit the Bloch representation of quantum
states and projection operations. A state $\rho$ acting on
$\mathbbm{C}^2$ can be expressed as
\beq
\rho=\frac{1}{2}(\mathbbm{1}_2+\boldsymbol{\rho}\cdot\boldsymbol{\sigma})
\eeq
 with $\boldsymbol{\rho}\in\mathbbm{R}^3$ being the so-called
Bloch vector of length $\|\boldsymbol{\rho}\|\leq 1$ and
$\boldsymbol{\sigma}=(\sigma_x,\sigma_y,\sigma_z)$ standing for a
vector consisting of the standard Pauli matrices\footnote{The Pauli matrices are defined as:
 $\sigma_{x}=\left(%
\begin{array}{cc}
  0 & 1 \\
  1 & 0 \\
\end{array}%
\right)$, $\sigma_{y}=\left(%
\begin{array}{cc}
  0 & -\mathrm{i} \\
  \mathrm{i} & 0 \\
\end{array}%
\right)$, $\sigma_{z}=\left(%
\begin{array}{cc}
  1 & 0 \\
  0 & -1 \\
\end{array}%
\right)$.}. Bloch vectors of unit length correspond to
rank-one projectors.

Within the Bloch representation the measurement operators $P_a$
associated to a generalized measurement (see the discussion above)
are represented by the Bloch vectors $\boldsymbol{\mathrm{p}}_a$ which due
to (\ref{norm2}) must satisfy
\begin{equation}
\sum_{a=0}^{k-1}\eta_a\boldsymbol{\mathrm{p}}_a=0.
\end{equation}
In particular, if $k=2$ the projections $P_a$ form a
a projective measurement and the above condition simplifies to
\begin{equation}
\boldsymbol{\mathrm{p}}_0+\boldsymbol{\mathrm{p}}_1=0.
\end{equation}
This implies that any two-outcome measurement defined on a qubit
Hilbert space is fully represented by a single vector, say
$\boldsymbol{\mathrm{p}}\equiv\boldsymbol{\mathrm{p}}_0$. Also, in
the case of two-outcome projective measurements we will mainly use
the more standard notation for the outcomes denoting them $\pm 1$
instead of $0,1$. Then, in the Bloch representation the projectors
corresponding to the outcomes can be represented as
\beq
P_{\pm}=\frac{1}{2}(\mathbbm{1}_2\pm\boldsymbol{\mathrm{p}}\cdot\boldsymbol{
\sigma } )
\eeq
and the mean value of an observable $\calA$ is just  $\langle
\calA\rangle=p(1|\calA)-p(-1|\calA)$.

\subsection{Bell-type experiment and local models}

Let us now consider $N$ spatially separated parties
$A^{(1)},\ldots,A^{(N)}=\boldsymbol{A}$ sharing some $N$-partite
quantum state $\rho_{\boldsymbol{A}}$ acting on a product Hilbert
space $\mathcal{H}=\mathbbm{C}^{d_1}\ot\ldots\ot\mathbbm{C}^{d_N}$
with $d_i<\infty$ $(i=1,\ldots,N)$ denoting the local dimensions.
On their share of $\rho_{\boldsymbol{A}}$, each party is allowed
to perform one of $m$ (possibly generalized) measurements
$\calA^{(i)}_{x_i}$ $(x_i=1,\ldots,m)$, each with $d$ outcomes,
which we enumerate as $a_i=0,1,\ldots,d-1$. We usually refer to
such a scenario as to $(N,m,d)$ scenario (of course, one can
consider a more general case of different number of measurements
and outcomes at each site, but such a scenario is a
straightforward generalization of the present one). In the case of
small number of parties $N=2,3$ we will rather denote them by
$A,B,C$, and the corresponding measurements and outcomes by
$\mathcal{A},\mathcal{B}, \mathcal{C}$ and $a,b,c$,
respectively.

The correlations generated in such an experiment
are described by a set of probabilities
\begin{eqnarray}\label{correlations}
p(a_1,\ldots,a_N|x_1,\ldots,x_N)&:=&
p(a_1,\ldots,a_N|\calA_{x_1}^{(1)},\ldots,\calA_{x_N}^{(N)})\non
&=&\tr \left[ \left( \calA_{x_1}^{(1)}\otimes \cdots\otimes
\calA_{x_N}^{(N)}\right) \rho_{\boldsymbol{A}}\right]
\end{eqnarray}
of obtaining results $a_1,\ldots,a_N$ upon measuring
$\calA_{x_1}^{(1)},\ldots,\calA_{x_N}^{(N)}$. In what follows, the
set (\ref{correlations}) will also be referred to as quantum
correlations or, simply, correlations. Also, it is useful to think
of the set $\{p(a_1,\ldots,a_N|x_1,\ldots,x_N)\}$, after some
ordering, as of a vector from $\mathbbm{R}^D$ with $D=d^Nm^N$.
Actually, by using the nonsignalling constraints (see, e.g.,
\cite{non-signalling,lifting}) one finds that a significantly
lower number $D'=[m(d-1)+1]^N-1$ of probabilities is sufficient to
fully describe the correlations (\ref{correlations}). For further
purposes let us finally notice that in the case $d=2$, one can
equivalently describe the correlations by a collection of
expectation values
\begin{equation}\label{correlators}
 \langle\mathcal{A}_{x_{i_1}}^{(i_1)}\ldots \mathcal{A}_{x_{i_k}}^{(i_k)}\rangle
\end{equation}
with $x_{i_1},\ldots,x_{i_k}=1,\ldots,m$,
$i_1<\ldots<i_k=1,\ldots,N$, and $k=1,\ldots,N$. Here
$\mathcal{A}_{x_{i_k}}^{(i_k)}$ are dichotomic observables with
outcomes $\pm1$. There are precisely $[m+1]^N-1$ of such
expectation values, which matches the number of independent
probabilities $D'$ in this scenario. This equivalence does not
hold if $d>2$ simply because the number of independent
probabilities exceeds that of correlators (\ref{correlators})
(see, nevertheless, Refs. \cite{Arnault,JRA} for possible ways to
overcome this problem).

It is known that the set of quantum correlations, denoted
$\mathbbm{Q}_{N,m,d}$, that can be produced in the above
experiment from all states and measurements is convex (provided
one does not constrain the dimension of the underlying Hilbert
space). A proper subset $\mathbbm{Q}_{N,m,d}$ is formed by those
correlations that the parties can obtain by using local strategies
and, as the only resource, some shared classical information,
often called \textit{shared randomness}, $\lambda$, distributed
among them with probability distribution $\omega(\lambda)$.
Mathematically, this is equivalent to saying that each probability
(\ref{correlations}) can be expressed as
\begin{eqnarray}\label{local}
\hspace{-2.2cm}
p(a_1,\ldots,a_N|\calA_{x_1}^{(1)},\ldots,\calA_{x_N}^{(N)})=\int_{\Omega}\mathrm{d}
\lambda\,\omega(\lambda)
p(a_1|\calA_{x_1}^{(1)},\lambda)\cdot\ldots\cdot
p(a_N|\calA_{x_N}^{(N)},\lambda)
\end{eqnarray}
with $\Omega$ denoting the set over which the random variable $\lambda$
is distributed and $p(a_i|\calA_{x_i}^{(i)},\lambda)$ are local
probability distributions conditioned additionally on $\lambda$.

It follows that in any scenario $(N,m,d)$, the set of local
correlations, i.e., those admitting the representation
(\ref{local}), is a polytope $\mathbbm{P}_{N,m,d}$ whose vertices
are the \textit{local deterministic correlations} of the form
\begin{equation}\label{deterministyczne}
p(a_1,\ldots,a_N|\calA_{x_1}^{(1)},\ldots,\calA_{x_N}^{(N)})=p(a_1|\calA_{x_1}^{(1)}
)\cdot\ldots\cdot p(a_N|\calA_{x_N}^{(N)})
\end{equation}
with the local probabilities being deterministic, that is,
$p(a_i|\calA_{x_i}^{(i)})\in\{0,1\}$ for all outcomes and
settings. In passing, let us notice that as $\mathbbm{P}_{N,m,d}$
has a finite number of vertices, the integration in (\ref{local})
can always be replaced by a finite sum over these vertices; for
further purposes it is however easier for us to use the integral
representation.

\begin{figure*}[h!]
\centering\includegraphics[clip,width=0.7\columnwidth]{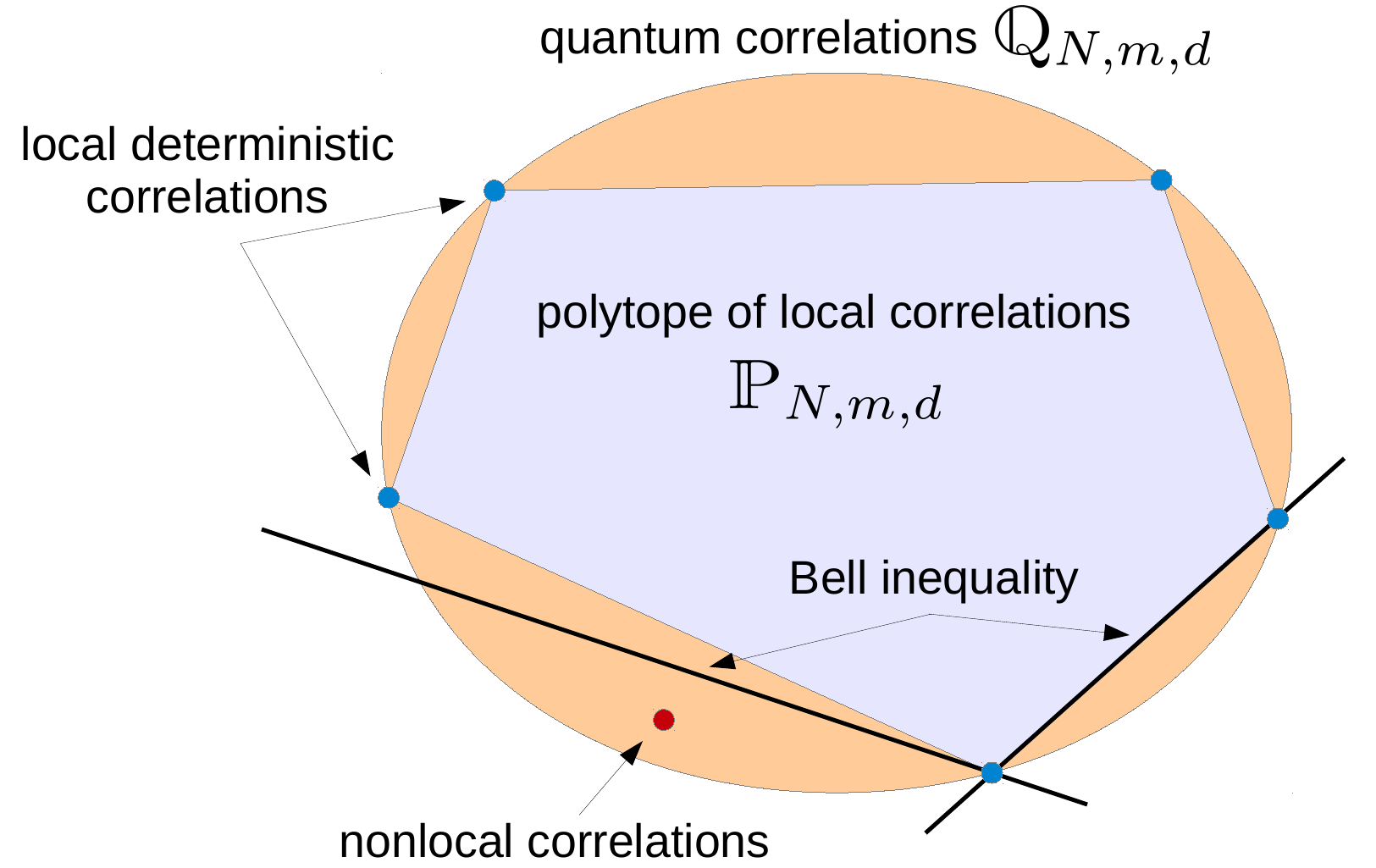}
\caption{\textbf{Quantum and local correlations.} Local
correlations form a polytope (blue gray region), denoted
$\mathbbm{P}_{N,m,d}$, with vertices being the local deterministic
correlations of the form (\ref{deterministyczne}) (blue points).
This set forms a proper subset of the set of quantum correlations,
$\mathbbm{Q}_{N,m,d}$, given by (\ref{correlations}) (orange
region). Correlations falling outside $\mathbbm{P}_{N,m,d}$ are
called nonlocal (red point) and can always be detected with the
help of a Bell inequality. On the scheme we give two examples of
such inequalities (thick lines), the one corresponding to the
facet of the polytope (on the right) is called \textit{tight},
and, accordingly, the other is said to be not
tight.}\label{fig:zbiory}
\end{figure*}

As already mentioned, $\mathbbm{Q}_{N,m,d}$ is strictly larger
than $\mathbbm{P}_{N,m,d}$ and quantum correlations that fall
outside $\mathbbm{P}_{N,m,d}$ are named \textit{nonlocal}. Natural
tools to witness \textit{nonlocality} in correlations are the
so--called {\it Bell inequalities} \cite{Bell} (see also
\cite{przegladowka-dani} and references therein for examples
thereof). These are linear inequalities that constrain the set of
local correlations and their violation is a signature of
nonlocality (see Fig. \ref{fig:zbiory}). For a given scenario
$(N,m,d)$, Bell inequalities can be most generally written as
\begin{equation}
 \beta:=\sum_{a_1,\ldots,a_N=0}^{d-1}\sum_{x_1,\ldots,x_N=1}^m
T_ {x_1,\ldots,x_N}^{a_1,\ldots,a_N}p(a_1,\ldots,a_N|x_1,\ldots,x_N)\leq
\beta_L,
\end{equation}
where $T$ is some tensor whose entries can always be taken nonnegative and
$\beta_L$ is the so-called classical bound of the inequality and is given
by $\beta_L=\max_{\mathbbm{P}_{N,m,d}}\beta$ (clearly, here it suffices to
maximize only over the vertices of $\mathbbm{P}_{N,m,d}$ in order to determine
$\beta_L$).

An illustrative example of a Bell inequality in the simplest
$(2,2,2)$ scenario is the famous Clauser-Horne-Shimony-Holt (CHSH)
Bell inequality \cite{CHSH}:
\begin{equation}\label{chsh-ineq}
\sum_{a,b=0}^1\sum_{x,y=1}^2p(a\oplus b=(x-1)(y-1)|xy)\leq 3,
\end{equation}
which can be restated in terms of the expectation values (\ref{correlators})
as
\begin{equation}\label{CHSHcor}
 |\langle \mathcal{A}_1\mathcal{B}_1\rangle
 +\langle \mathcal{A}_1\mathcal{B}_2\rangle+\langle
\mathcal{A}_2\mathcal{B}_1\rangle-\langle
\mathcal{A}_2\mathcal{B}_2\rangle|\leq 2.
\end{equation}
Here $\mathcal{A}_i$ and $\mathcal{B}_i$ $(i=1,2)$ are pairs of
dichotomic observables with eigenvalue $\pm 1$ measured by the
parties $A$ and $B$, respectively. It is known that in the
$(2,2,2)$ scenario the CHSH inequality (\ref{CHSHcor}) is the only
``relevant'' Bell inequality as it defines all facets of
$\mathbbm{P}_{2,2,2}$ \cite{Fine}. In other words, if a bipartite
state $\rho_{AB}$ does not violate the inequality (\ref{CHSHcor})
for any choice of the measurements $\mathcal{A}_i$ and
$\mathcal{B}_i$, then it does not violate any other Bell
inequality in this scenario. The CHSH inequality is an example of
a {\it correlation Bell inequality}, in which only the joint terms
$\langle \mathcal{A}_i\mathcal{B}_j\rangle$ appear.

A schematic depiction of all the above concepts is provided in Fig. \ref{fig:zbiory}.

Now, within this framework, we call the state $\rho$ \textit{local
in the scenario $(N,m,d)$} if the probability distribution that
can be generated from it is local for any choice of the
measurements $\calA_{x_i}^{(i)}$ with $x_i=1,\ldots,m$ and
$i=1,\ldots,N$. Notice that if $\rho$ is local in the scenario
$(N,m,d)$, then it is local in any  scenario $(N,m',d')$ with
either $m'<m$ or $d'<d$. On the other hand, there are states that
are local in some scenario $(N,m,d)$, but their nonlocality can be
revealed only when the number of measurements or outcomes is
increased.

Finally, a state $\rho$ is called \textit{local} if it is local in
a scenario with any number of measurement and outcomes.
Equivalently, one says that $\rho$ has a \textit{local-hidden
variable (LHV) model}  or simply a \textit{local model}. Notice
that in such  case, one can drop all the subscripts $x_i$ in
(\ref{local}) because any probability generated from $\rho$ must
take such a form. As a result, (\ref{local}) simplifies to
\begin{eqnarray}\label{local-model}
\fl
p(a_1,\ldots,a_N|\calA^{(1)},\ldots,\calA^{(N)})=\int_{\Omega}\mathrm{d}
\lambda\,\omega(\lambda)p(a_1|\calA^{(1)},\lambda)\cdot\ldots\cdot
p(a_N|\calA^{(N)},\lambda).
\end{eqnarray}

On the other hand, if there is a scenario $(N,m,d)$ in which for some
choice of measurements the resulting probability distribution
(\ref{correlations}) does not admit the form (\ref{local}), we call $\rho$
\textit{nonlocal}.

Of course, one can further distinguish the case when $\rho$ is
local only under projective measurements, i.e., (\ref{local})
holds if the measurements $\calA^{(i)}$ with $i=1,\ldots,N$ are
projective. Then, one says that $\rho$ has a local model for
projective measurements. Moreover, for $d=2$, one can also formulate
the definition of locality of quantum states
in terms of expectation values (\ref{correlators}). Then, every expectation
value involving more than two parties admits the form (\ref{local-model}) with
local probabilities replaced by the corresponding one-body mean values
``conditioned'' on $\lambda$.

Notice furthermore that a model reproducing
global probability distribution automatically reproduces local statistics as the
latter are just the marginals of the former. The situation is more subtle when
we work with the expectation values (\ref{correlators}).
Here, a local model realizing expectation values involving some number of
parties, say $k$, does not necessarily properly describes expectation values
involving less than $k$ parties. In particular, in the bipartite case,
a model
reproducing joint expectation values $\langle \mathcal{A}\mathcal{B}\rangle$
may not reproduce $\langle \calA\rangle$ and $\langle \calB\rangle$; we refer
to such model as to a model for joint correlations. This observation will be of
particular importance in Sec. \ref{groth}.

\begin{figure*}[h!]
\centering\includegraphics[clip,width=0.4\columnwidth]{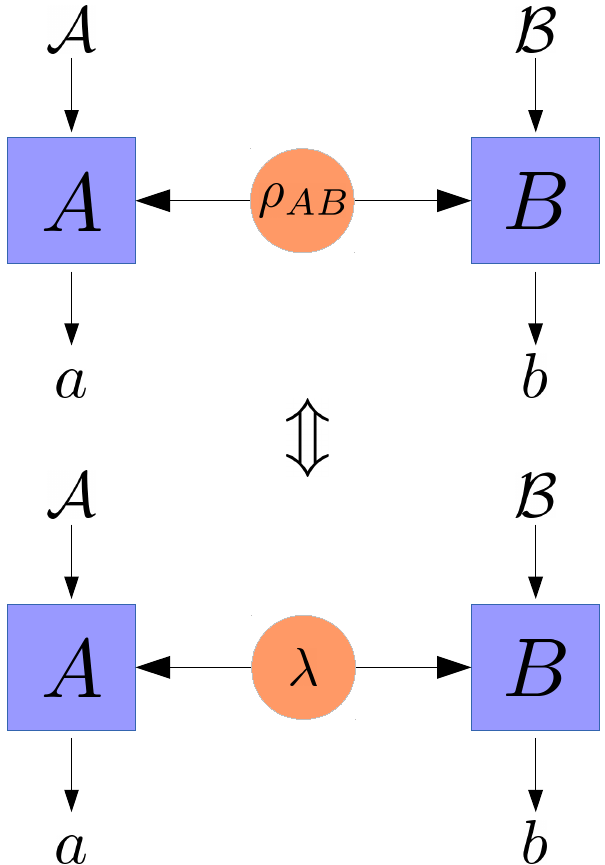}
\caption{\textbf{Simulating quantum probabilities with classical information.} A schematic
illustration of the concept of locality of quantum states. For simplicity we
consider the bipartite case. Imagine that the parties $A$ and $B$ perform a pair
of measurements $\mathcal{A}$ and $\mathcal{B}$ on some bipartite (entangled)
state $\rho_{AB}$ obtaining results $a$ and $b$, respectively. Locality of
$\rho_{AB}$ means that the resulting probability distribution
$p(a,b|\mathcal{A},\mathcal{B})$ for any pair $\mathcal{A},\mathcal{B}$ can also
be obtained if the parties shared some classical information represented by
$\lambda$ instead of $\rho_{AB}$. Phrasing differently, one can simulate
statistics arising from local quantum states by using purely
classical resources.}\label{fig:resources}
\end{figure*}

From the point of view of the resources shared by the parties, the
question of whether a state has a local model is the question of
whether  shared classical randomness can replace it in the
described setup (see Fig. \ref{fig:resources}).

It is clear that an unentangled state is trivially local in any
scenario as the global probability of local measurements is a sum
of factorized components. In the remainder of this paper we thus
focus on providing LHV models for entangled states and genuinely multipartite
entangled state in the multipartite case (cf. Sec. \ref{multiparty}). In
the most general setting such a model simulating statistics of the
measurements $\calA$ and $\calB$ with the results $a$ and $b$,
respectively, will be given as follows (for brevity we give a
bipartite version):

\linka
\begin{center}
\textbf{General local model}
\end{center}
\begin{itemize}
\item[(0)]Alice and Bob are distributed some classical information
$\lambda$ belonging to some $\Omega$ with probability distribution
$\omega(\lambda)$,
\item[(1)] Alice outputs $a$ according to some probability
distribution $p(a|\calA,\lambda)$ conditioned on $\mathcal{A}$ and
$\lambda$,
\item[(2)] and, likewise, Bob outputs $b$ according to
the probability distribution $p(b|\calB,\lambda)$.
\end{itemize}
\linka

In the above, $p(a|\calA,\lambda)$ and $p(b|\calB,\lambda)$, also
called {\it response functions}, determine the strategies parties
need to employ to simulate measurements on a given quantum state
(it is not difficult to see that without loss of generality they could be taken
deterministic \cite{Fine}). Once these mappings are proposed it will be the main
task to verify whether (\ref{local-model}) holds with l.h.s.
arising from measurements on a quantum state of interest.

Notice that in the general model we stated, the nature of
$\lambda$ is not an issue. It can be a single variable (discrete
or continuous), a set of variables, etc. Also the case of Alice
and Bob receiving different $\lambda$s is covered in this very
general formulation.

\section{Bipartite quantum states}

We start our tour through local models with bipartite scenarios.
In such setting the first and most famous model is due to Werner
\cite{Werner} who already in 1989 realized that projective
measurements on some entangled states give rise to local
statistics. These states belong to the class which is
usually referred to as the Werner states.
Besides presenting the original model we will also discuss several
of its modifications. Many years later the result by Werner was generalized by
Barrett \cite{Barrett} who proved that local models may account for
statistics even of generalized measurements on some entangled
Werner states. The price one had to pay for such extension was
the diminishment of the region of parameters of the applicability
of the model. At this moment, it is not known whether Werner
states are local for projective and generalized measurements in
the same region of the parameter.

An important line of research in the domain has been the analysis
of the nonlocal properties of noisy states, i.e., the ones which
arise as the mixture of some state with the white (completely
depolarized) noise. This problem was addressed by \acin \etal
\cite{Acin-grod} and Almeida \etal \cite{Almeida}. In the former
work, nonlocal properties of such states were related to
the Grothendieck constant \cite{Grothendieck}, establishing, in
particular, the region in which two-qubit Werner state is local
for projective measurement. In the latter, building on the
previous works by Werner and Barrett about local models and the
protocol by Nielsen for the deterministic conversion between pure entangled  states
by local operations and classical communication (LOCC) \cite{Nielsen},
values of the critical noise thresholds for locality of noisy states were obtained.

We will conclude this section with a recent result by Hirsch \etal
\cite{Hirsch} who found a method of constructing states with local
models for POVMs from states with local models only for two
outcome projective measurements.

\subsection{Werner's model}\label{werner-model}

The class of states considered by Werner consists of all
two-qudit states that are invariant under a bilateral action of any unitary
operation, i.e., they commute with $U\ot U$ for any unitary $U$ acting on
$\mathbbm{C}^d$. Such states are of the form
\begin{eqnarray}\label{WernerState}
\rho_{\mathrm{W}}(d,p)&=&p
\frac{2P_{d}^{(-)}}{d(d-1)}+(1-p)\frac{\mathbbm{1}_d}{d^2}
\nonumber\\
&=&\frac{1}{d(d-1)}\left(\frac{d-1+p}{d}\mathbbm{1}_d-pV_d\right),
\end{eqnarray}
where $P_{d}^{(-)}$ ($P_d^{(+)}$) stands for the projector onto
the antisymmetric (symmetric) subspace of
$\mathbbm{C}^d\otimes\mathbbm{C}^d$
and $V_d$ denotes the so-called swap
operator defined through
\begin{equation}
V_d\ket{\phi_1}\ket{\phi_2}=\ket{\phi_2}\ket{\phi_1}
\end{equation}
for any pair $\ket{\phi_1},\ket{\phi_2}\in\mathbbm{C}^d$. It is
useful to recall that $P_{d}^{(\pm)}=(\mathbbm{1}_d\pm V_d)/2$.
Also, when $d=2$, $P_{2}^{(-)}=\proj{\psi_-}$ is just the
projection onto the two-qubit singlet state
\beq \ket{\psi_-}=1/\sqrt{2} (\ket{01}-\ket{10}). \eeq

It was shown in Ref. \cite{Werner} that Werner states
$\rho_{\mathrm{W}}(d,p)$ are separable if and only if $p\leq
p_{\mathrm{sep}}^{W,c}$ with
\beq p_{\mathrm{sep}}^{W,c}\equiv \frac{1}{d+1}. \eeq

In particular, it is very easy to see that they are entangled for any $p>1(d+1)$
as the swap operator $V_d$ is also an entanglement witness
\cite{horodeccy-witness,terhal-witness} detecting them, i.e.,
$\Tr[\rho_{\mathrm{W}}(d,p)V_d]<0$ in this region.

Let us now show, following Ref. \cite{Werner}, that for any $p\le
p_{\mathrm{PM}}^{\mathrm{W},\mathrm{c}}$ with
\beq p_{\mathrm{PM}}^{\mathrm{W,c}}\equiv \frac{d-1}{d} \eeq
 these states always give rise
to local statistics when measured with projective measurements. In
fact, it is enough to show this for the boundary value of the
mixing parameter $p=p_{\mathrm{PM}}^{\mathrm{W,c}}$, for which Eq.
(\ref{WernerState}) reduces to
\begin{equation}\label{WernerState2}
\rho_{\mathrm{W}}^d\left(d,\textstyle{\frac{d-1}{d}}\right)=\displaystyle
\frac{1}{d^2} \left(\frac{d+1}{d}\mathbbm{1} _d-V_d\right).
\end{equation}
For all values of $p<p_{\mathrm{PM}}^{\mathrm{W,c}}$ the result
will follow immediately because one can always obtain the
corresponding $\rho_{\mathrm{W}}(d,p)$ by mixing
$\rho_{\mathrm{W}}(d,p_{\mathrm{PM}}^{\mathrm{W,c}})$ with some
portion of the white noise $\mathbbm{1}_d/d^2$ which itself
clearly admits a local model and the state given by the mixture of states with local models
has also a local model.

Now, it directly stems from Eq. (\ref{WernerState2}) that the
´´quantum''  probability of obtaining outcomes $a$ and $b$ when
performing the von Neumann measurements $\calA$ and $\calB$
represented by projections $\{P_a\}$ and $\{Q_b\}$, respectively,
amounts to
\begin{equation}\label{ProbsWerner}
p_Q^W(a,b|\calA,\calB)=\frac{1}{d^2}\left[\frac{d+1}{d}-\Tr(P_aQ_b)\right],
\end{equation}
where we have used the well-known property of the swap operator that
$\Tr(V_dX\otimes Y)=\Tr(XY)$ for any pair of matrices $X,Y.$

To simulate (\ref{ProbsWerner}) with local strategies Werner
proposed the following model:

\linka
\begin{center}
\textbf{LHV model for projective measurements on Werner
states (Werner's model) \cite{Werner}}
\end{center}
\begin{itemize}
\item[(0)] Alice and Bob are distributed shared randomness represented by a
one-qudit pure state
$\ket{\lambda}\in\mathbbm{C}^d$ (the hidden state space is
$\Omega_d=\{\ket{\lambda}\in\mathbbm{C}^d\,|\,\langle\lambda|\lambda\rangle=1\}
$) with the unique distribution invariant under any unitary
operation $\omega_d(\lambda)$,
\item[(1)] Alice returns the outcome
$a$ for which the overlap $\langle\lambda|P_a|\lambda\rangle$ is
the smallest one, i.e., her response function reads
\begin{equation}\label{ProbAA}
p(a|\calA,\lambda)=\left\{
\begin{array}{ll}
1, \quad \mathrm{if}\;
\langle\lambda|P_a|\lambda\rangle=\displaystyle\min_\alpha\langle\lambda|P_{
\alpha }
|\lambda\rangle\\
0,\quad\mathrm{otherwise}
\end{array},
\right.
\end{equation}
 \item[(2)] Bob returns the outcome $b$ with probability
\begin{equation}\label{ProbB}
p(b|\calB,\lambda)=\langle\lambda|Q_b|\lambda\rangle.
\end{equation}
\end{itemize}
\linka

For further purposes it is worth noting that the response
functions (\ref{ProbAA}) and (\ref{ProbB}) have the following
equivariance property
\begin{equation}
 p(a|\calA,U\lambda)=p(a|U\calA U^{\dagger},\lambda),\quad
p(b|\calB,U\lambda)=p(b|U\calB U^{\dagger},\lambda)
\end{equation}
for any unitary $U$ acting on $\mathbbm{C}^d$. Let us mention in
passing that originally in Ref. \cite{Werner} both the response
functions were exchanged. This, however, does not affect the final
result as the Werner states are permutationally invariant. The
latter also means that we can even symmetrize the roles of the
parties in the protocol. This can be easily done by augmenting the
protocol with an additional two--valued hidden variable
determining which type of the function, (\ref{ProbAA}) or
(\ref{ProbB}), each of the party must choose as the response.

 Having all the necessary ingredients, we can now
demonstrate that indeed $p_Q^W(a,b|\calA,\calB)$ admits the form
(\ref{local-model}) with $N=2$, i.e.,
\begin{equation}\label{integralWerner}
p_Q^W(a,b|\calA,\calB)
=\int_{\Omega_d}\mathrm{d}\lambda\,
\omega_d(\lambda)p(a|\calA,\lambda)p(b|\calB,\lambda).
\end{equation}
To be more instructive and explanatory, we will exploit for this
purpose the approach of Ref. \cite{Mermin} rather than the
original one by Werner. We start by noting that the value of the
integral in (\ref{integralWerner}) is, by the very construction of
the model, invariant under any unitary rotation applied to
$\ket{\lambda}$. This means that we can decompose $\ket{\lambda}$
in the eigenbasis $\{\ket{i}\}$ of Alice's observable $\calA$,
i.e.,
\begin{equation}\label{decomposition}
\ket{\lambda}=\sum_{i}\lambda_i\ket{i}
\end{equation}
with $\lambda_i$ being some complex coefficients such that
$|\lambda_1|^2+\ldots+|\lambda_d|^2=1$. In what follows we will
exploit their polar representation, that is,
$\lambda_i=r_i\mathrm{exp}(\mathrm{i}\phi_i)$ with $0\leq r_i\leq
1$ and $\phi_i\in[0,2\pi)$ for any $i$. Clearly,
$r_1^2+\ldots+r_d^2=1$. Additionally, without any loss of
generality, we can assume that the projection corresponding to the
particular outcome $a$ appearing in (\ref{integralWerner}) is
simply $P_a=\proj{1}$. All this allows us to rewrite the
probabilities (\ref{ProbAA}) and (\ref{ProbB}) as
\begin{equation}\label{Manchego}
p(a|\calA,\lambda)=
\left\{
\begin{array}{ll}
1,\quad r_1^2=\displaystyle\min_{\alpha=2,\ldots,d}r_{\alpha}^2\\
0,\quad\mathrm{otherwise}
\end{array}
\right.
\end{equation}
and
\begin{eqnarray}
p(b|\calB,\lambda)&=&\sum_{i,j=1}^d\langle i|Q_b|j\rangle r_ir_j\mathrm{e}^{\mathrm{i}(\phi_j-\phi_i)}\nonumber\\
&=&\sum_{i=1}^d\langle i|Q_b|i\rangle r_i^2+\sum_{i\neq
j}^d\langle i|Q_b|j\rangle
r_ir_j\mathrm{e}^{\mathrm{i}(\phi_j-\phi_i)}
\end{eqnarray}
respectively. Inserting these into (\ref{integralWerner}) and performing
integrations over all the angles
$\phi_i$ (clearly, the second summand in the above formula gives zero), one
arrives at
\begin{eqnarray}\label{Castelldefels}
\int_{{\Omega}_d}\mathrm{d}\lambda
\,\omega_d(\lambda)p(a|\calA,\lambda)p(b|\calB,\lambda)=\frac{1}{N}\sum_{i=1}^{d
} \langle i|Q_b|i\rangle J(u_i),
\end{eqnarray}
with the functional
\begin{eqnarray}\label{integrals}
J(h)=\int_0^1\mathrm{d}u_1\int_{u_1}^1\mathrm{d}u_2\ldots
\int_{u_1}^1\mathrm{d}u_d h(u_1,\ldots,u_d) \delta(u_1+\ldots
+u_d-1),\non
\end{eqnarray}
where we have made the substitution $r_i^2=u_i$. The Dirac delta in the above is
due to the normalization of $\ket{\lambda}$, while the lower bounds in the
integrations over $u_2,\ldots,u_d$ follow from the condition in
(\ref{Manchego}). Finally,
\begin{equation}\label{N}
N=\int_0^1\mathrm{d}u_1\ldots \int_0^1\mathrm{d}u_d \delta(u_1+\ldots
+u_d-1).
\end{equation}

One can significantly simplify the computation of the right-hand
side of (\ref{Castelldefels}) by noting that the integrals in
(\ref{integrals}) are manifestly symmetric under any permutation
of the variables $u_2,\ldots,u_d$ (but not $u_1$!) and therefore
$J(u_2)=\ldots=J(u_d)$. Moreover, the delta function in
(\ref{integrals}) allows us to conclude that
$J(u_2)+\ldots+J(u_d)=J(1)-J(u_1)$, which in turn means that
$J(u_i)=[J(1)-J(u_1)]/(d-1)$ for any $i=2,\ldots,d$. Putting
pieces together, we have that
\begin{eqnarray}\label{Castelldefels2}
\fl\int_{\Omega_d}\mathrm{d}\lambda\,
\omega_d(\lambda)p(a|\calA,\lambda)p(b|\calB,\lambda)=\frac{[d\Tr(P_aQ_b)-1]
J(u_1)+
[1-\Tr(P_aQ_b)]J(1)}{N(d-1)},
\end{eqnarray}
where we have substituted back $P_a$ for $\proj{1}$, which in
particular means that $\langle1|Q_b|1\rangle$ is now
$\Tr(P_aQ_b)$, and we have also used the fact that $\Tr Q_b=1$ for
any $b$. To complete the proof, one then needs to determine the
values of the integrals $J(1)$ and $J(u_1)$.
Computation of the latter has been done in \ref{appendix} and its
value has been found to be
\beq\label{jot-u-jeden} J(u_1)=N/d^3. \eeq
 The value of $J(1)$ can be inferred directly from
the above formula (\ref{Castelldefels2}): exploiting the fact that
$p(a|\calA,\lambda)$ and $p(b|\calB,\lambda)$ are proper
probability distributions, by summing both its sides over $a$ and
$b$, one obtains that $J(1)=N/d$. Inserting both integrals into
the right-hand side of Eq. (\ref{Castelldefels2}), one finally
obtains the quantum probability (\ref{ProbsWerner}).

As a result, the above model allows one to show that the
probabilities arising from projective measurements performed on
the subsystems of the Werner states can always be simulated by a
local model for any $p\leq (d-1)/d$. Since the boundary value
$(d-1)/d$ is larger than the separability threshold $1/(d+1)$ of
$\rho_{\mathrm{W}}(d,p)$, this means that for any $d$ there are
entangled states with local model for projective measurements.
Noticeably, the critical value grows with $d$, while the one for
separability drops, meaning that the models becomes more powerful
for larger $d$ (cf. Fig. \ref{fig:Werner}).

\begin{rem}\label{remark-werner} Let us also
notice that the above model, initially designed for two-outcome
projective measurements at each site, works also if a generalized
measurement with any number of outcomes is allowed at Bob's site.
This directly follows from the facts that Bob's response function
is linear in the measurement operators and that the measurement
operators of a POVM $\mathcal{B}$ can be taken as
$\mathcal{B}_b=\xi_bQ_b$ with rank-one operators $Q_b$ and some
positive constants $\xi_b$ (see Sec. \ref{Sec:Measurement}).
Therefore the probabilities realized by the model in this case are
$p_{\mathrm{L}}(a,b|\calA,\calB)= \xi_b
p_{\mathrm{L}}(a,b|\calA,\calB')$, where
$p_{\mathrm{L}}(a,b|\calA,\calB')$ denote unnormalized
``probabilities'' obtained by measuring projectors $Q_b$ on Bob's
site. At the same time, one realizes that the quantum
probabilities are $p_{\mathrm{Q}}(a,b|\calA,\calB)=\xi_b
p_{\mathrm{Q}}(a,b|\calA,\calB')$, and therefore
$p_{\mathrm{L}}(a,b|\calA,\calB)=p_{\mathrm{Q}}(a,b|\calA,\calB)$.
\end{rem}

\begin{rem}\label{rem1}
Interestingly, as shown by Gisin and Gisin in Ref.
\cite{GisinGisin}, one can simulate the statistics of the singlet
state (or, equivalently,
 the Werner state for $p=1$) with a local model if Alice or Bob (or both) are allowed not to provide outcomes.
To state the model we now switch, for simplicity, to the Bloch representation, however, all that follows can be
 restated in terms of vectors from $\mathbbm{C}^d$.

Assume that Alice and Bob receive unit vectors $\boldsymbol{\mathrm{a}}$
and $\boldsymbol{\mathrm{b}}$ representing local projective measurements
$\calA$ and $\calB$, respectively, and a unit vector
$\boldsymbol{\lambda}$ from the Bloch sphere generated according
to the uniform distribution
$\widetilde{\omega}(\boldsymbol{\lambda})=1/4\pi$. Now, Alice
follows the same strategy as in Werner's model, that is, she
always outputs $-\mathrm{sgn}(\boldsymbol{\mathrm{a}}\cdot\boldsymbol{\lambda})$
with $\mathrm{sgn}$ denoting the sign function.
Then, with probability $|\boldsymbol{\mathrm{b}}\cdot\boldsymbol{\lambda}|$,
Bob accepts $\boldsymbol{\lambda}$ and outputs
$\mathrm{sgn}(\boldsymbol{\mathrm{b}}\cdot\boldsymbol{\lambda})$ in this
case, while if he does not accept $\boldsymbol{\lambda}$, he does not
give any outcome. The local expectation values $\langle
\calA\rangle=\langle \calB\rangle=0$, while the correlation
produced by this model taken over all the ``accepted"
$\boldsymbol{\lambda}$ are given by
\begin{eqnarray}\label{lookat}
\langle \calA\otimes\calB\rangle&=&-\frac{1}{2\pi}
\int\mathrm{d}\boldsymbol{\lambda}|\boldsymbol{\mathrm{b}}\cdot\boldsymbol{
\lambda}|\mathrm{sgn}(\boldsymbol{\mathrm{a}}\cdot\boldsymbol{\lambda})
\mathrm{sgn}(\boldsymbol{\mathrm{b}}\cdot\boldsymbol{\lambda})\nonumber\\
&=&-\frac{1}{2\pi} \int\mathrm{d}\boldsymbol{\lambda}\,
\mathrm{sgn}(\boldsymbol{\mathrm{a}}\cdot\boldsymbol{\lambda})\,(\boldsymbol{
\mathrm{b}} \cdot\boldsymbol{\lambda})\nonumber\\
&=&-\boldsymbol{\mathrm{a}}\cdot\boldsymbol{\mathrm{b}},
\end{eqnarray}
where to obtain the last equation we have followed the calculation
from \ref{appendix}. The corresponding probabilities are given by
\begin{equation}
p(a,b|\calA,\calB)=\frac{1}{4}(1-ab\,\boldsymbol{\mathrm{a}}\cdot\boldsymbol{
\mathrm{b}})\qquad
(a,b=\pm 1)
\end{equation}
and they agree with (\ref{ProbsWerner}) for $d=2$.

Looking at Eq. (\ref{lookat}), one can realize that this model can
also be understood as one with a distribution
$(1/2\pi)|\boldsymbol{\mathrm{b}}\cdot\boldsymbol{\lambda}|$ which
depends on the measurement of Bob. In other words, a model with a
distribution that depends on the measurements of one of the
parties can reproduce even highly nonlocal correlations.
In fact, this observation was used in
\cite{degorre} to construct models for the simulation of entangled
states using classical communication or supra-quantum resources.
\end{rem}

\begin{rem} Building on the above result,
one can introduce a local model reproducing
statistics obtained on a two-qubit state that differs from the
two-qubit Werner state \cite{Hirsch}. Alice follows the same strategy as above,
while Bob whenever he does not accept $\boldsymbol{\lambda}$,
outputs $b=\pm 1$ with probability $(1\pm \langle
\eta|\boldsymbol{\mathrm{b}}\cdot\boldsymbol{\sigma}|\eta\rangle)/2$. After
some calculation one finds that the class of states whose correlations are
simulated by this model is given by
\begin{equation}\label{Raimat}
\rho=p\proj{\psi_-}+(1-p)\frac{\mathbbm{1}_2}{2}\ot
\proj{\eta}.
\end{equation}
with $p\leq 1/2$.
\end{rem}

\begin{rem}\label{rem3}
Let us conclude by noting that one can also extend the
applicability of Werner's model by playing with the distribution
$\omega$. An example of a distribution different than the uniform
one has been recently given  in \cite{BVQB} and reads
$\omega(\boldsymbol{\lambda})=(1+\lambda_3)/4\pi$, where
$\lambda_3$ is the third coordinate of the Bloch vector
$\boldsymbol{\lambda}$. In such case, the model corresponds to the
two-qubit state
\begin{equation}
\rho(p)=p
\proj{\psi_-}+\frac{1-p}{5}\left(2\proj{0}\ot\frac{\mathbbm{1}_2}{2}+3\frac{\mathbbm
{ 1 } _2 } {2}\ot\proj{1} \right)
\end{equation}
with $p\leq 1/2$. Notice that this state is a convex combination of two states
given in Eq. (\ref{Raimat}).
\end{rem}

\subsection{Barrett's model for the Werner states}\label{baret}

Let us now present the local model for general measurements for
the Werner states due to Barrett \cite{Barrett}. Now, Alice and
Bob want to simulate the probability distribution arising from
some generalized measurements, $\calA$ and $\calB$, with the
measurement operators $\calA_a=\eta_aP_a$ and $\calB_b=\xi_bQ_b$ (cf.
Sec. \ref{Sec:Measurement}) performed on
Werner states (\ref{WernerState}). One quickly finds that this
distribution is not much different from that for projective
measurements (\ref{ProbsWerner}) and  simply reads
\begin{equation}\label{ProbsWerner2}
 p_Q^W(a,b|\calA,\calB)=\frac{\eta_a\xi_b}{d(d-1)}\left[\frac{d-1+p}{d}
-p\Tr(P_aQ_b)\right].
\end{equation}
The model now goes as follows:

\linka
\begin{center}
\textbf{LHV model for POVMs on Werner states  (Barrett's
model) \cite{Werner}}
\end{center}
\begin{itemize}
\item[(0)] As in Werner's model, the set of hidden states and the distribution
are, respectively, $\Omega_d$ and $\omega_d$,
\item[(1)] Alice returns the outcome
$a$ according to the response function:
\begin{eqnarray}\label{respuestaA}
p(a|\calA,
\lambda)&=&\langle\lambda|\calA_a|\lambda\rangle\Theta(\langle\lambda|P_a|\lambda
\rangle-1/d)\nonumber\\
&&+\left(1-\sum_{\alpha}\langle\lambda|\calA_{\alpha}
|\lambda\rangle\Theta(\langle\lambda|P_a|\lambda
\rangle-1/d)\right)\frac{\eta_a}{2},
\end{eqnarray}
 \item[(2)] Bob returns the outcome $b$ with the probability
\begin{equation}\label{respuestaB}
p(b|\calB,\lambda)=\frac{\xi_b}{d-1}(1-\langle\lambda|Q_b|\lambda\rangle).
\end{equation}
\end{itemize}
\linka

In the above, $\Theta$ is the Heaviside step function, i.e.,
$\Theta(x)=1$ for any $x\geq 0$ and $\Theta(x)=0$ for $x<0$.

\begin{figure*}[h!]
\centering\includegraphics[clip,width=0.5\columnwidth]{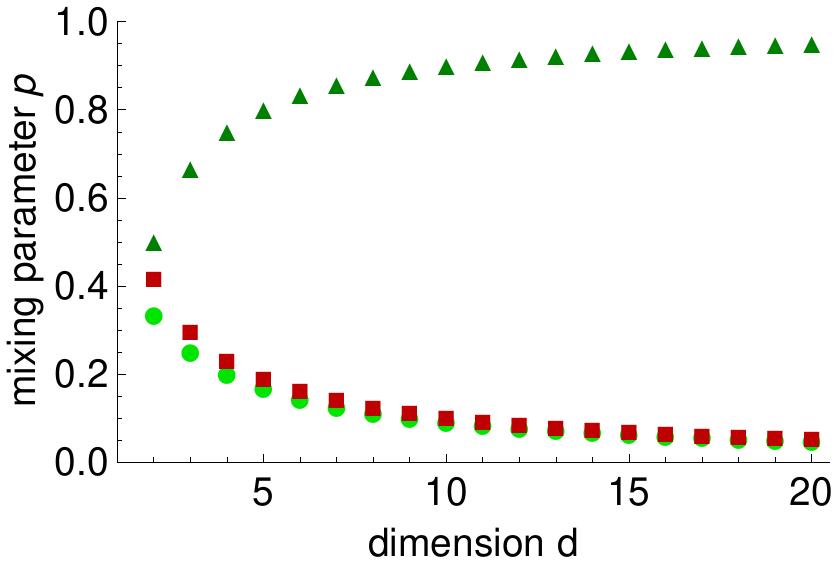}
\caption{\textbf{Critical probabilities for the Werner states.}
Comparison of three critical values of the probabilites for the
Werner states for $2\leq d\leq 20$: $p^{W,c}_{\mathrm{sep}}$
(green dots), $p^{W,c}_{\mathrm{POVM}}$ (red squares),
$p^{W,c}_{\mathrm{PM}}$ (green triangles). Noticeably,
while $p_{\mathrm{sep}}^{W,c}$ decays with $d$, the critical value
$p^{W,c}_{\mathrm{PM}}$ grows, implying that for large $d$ almost
all the entangled Werner states are simulable by local
models.}\label{fig:Werner}
\end{figure*}

To demonstrate that the above model recovers (\ref{ProbsWerner2})
and also to determine the range of $p$ for which this is the case,
let us insert the response functions (\ref{respuestaA}) and
(\ref{respuestaB}) into (\ref{integralWerner}). After some
algebra, this leads us to
\begin{eqnarray}\label{PL}
p_{L}^W(a,b|\calA,\calB)&=&\frac{\eta_a\xi_b}{d^2}-\frac{1}{d-1}J_{ab}+\frac{
\xi_b } { d-1
}\sum_ { \beta }J_{a\beta}\nonumber\\
&&+\frac{\eta_a}{d(d-1)}\sum_{\alpha}J_{\alpha
b}-\frac{\eta_a\xi_b}{d(d-1)}\sum_{\alpha\beta}J_{\alpha\beta}
\end{eqnarray}
with
\begin{equation}\label{Jab}
J_{ab}=\eta_a\xi_b\int_{\Omega_d}\mathrm{d}\lambda
\,
\omega_d(\lambda)\Theta(\langle\lambda|P_a|\lambda\rangle-1/d)\langle\lambda|P_
a |
\lambda\rangle\langle\lambda|Q_b|
\lambda\rangle.
\end{equation}
In order to perform the above integration, let us decompose $\ket{\lambda}$ as in (\ref{decomposition})
 with $\lambda_i=r_i\mathrm{exp}(\mathrm{i}\phi_i)$ and the basis $\{\ket{i}\}$
chosen so that it contains a vector corresponding to $P_a$ (recall
that now the projectors $P_a$ as well as $Q_b$ do not have to be
orthogonal). As previously, we can assume that $P_a=\proj{1}$.
With this parametrization, one finds that (\ref{Jab}) becomes now
\begin{equation}\label{Jab2}
J_{ab}=\frac{\eta_a\xi_b}{N}\sum_{k=1}^d\langle\lambda|Q_b|\lambda\rangle
\widetilde{J}[u_1u_k],
\end{equation}
where the normalization factor $N$ is given by (\ref{N}),
while the functional $\widetilde{J}[h]$ is defined as
\begin{eqnarray}
 \widetilde{J}[h]=\int_{1/d}^1\mathrm{d}u_1\int_{0}^1\mathrm{d}u_2\ldots
\int_{0}^1\mathrm{d}u_d\,h(u_1,\ldots,u_d)\,\delta(u_1+\ldots+u_d-1).\nonumber\\
\end{eqnarray}
By noting that
$\widetilde{J}[u_1u_2]=\ldots=\widetilde{J}[u_1u_d]$, one easily
sees that
$\widetilde{J}[u_1u_k]=\{\widetilde{J}[u_1]-\widetilde{J}[u_1^2]\}/(d-1)$,
which allows one to further simplify (\ref{Jab2}) to
\begin{eqnarray}
J_{ab}=\frac{\eta_a\xi_b}{N(d-1)}\left\{[d\Tr(P_aQ_b)-1]\widetilde{J}
[u_1^2]+[1-\Tr(P_aQ_b)] \widetilde{J}[u_1]\right\}.\nonumber\\
\end{eqnarray}
After inserting this into (\ref{PL}), one arrives at the following
expression
\begin{equation}
p^W_{L}(a,b|\calA,\calB)=\eta_a\xi_b\left\{\frac{1}{d^2}
+\frac{d\widetilde{J}[u_1^2]-\widetilde{J}
[u_1]}{N(d-1)^2}\left[\frac{1}{d}-\Tr(P_aQ_b)\right]\right\},
\end{equation}
which, when compared with (\ref{ProbsWerner2}), shows that the
above model reproduces joint probabilities for the Werner states
for any $p\leq p^{W,c}_{\mathrm{POVM}}$ with the critical
probability given by
\begin{equation}
p^{W,c}_{\mathrm{POVM}}=d\frac{d\widetilde{J}[u_1^2]-\widetilde{J}[u_1]}{N(d-1)}
.
\end{equation}
To finally determine the explicit value of
$p_{\mathrm{POVM}}^{W,c}$ one has to compute the integrals
$\widetilde{J}(u_1)$ and $\widetilde{J}(u_1^2)$. This can be done
by exploiting basically the same approach as in the case of
projective measurements (the detailed calculations are moved to \ref{appendix}),
which leads us to
\begin{equation}\label{jot-u-jeden-tylda}
\widetilde{J}[u_1]=\frac{N}{d}\left(\frac{
2d-1}{d}\right)\left(\frac{d-1}{d}\right)^{d-1}
\end{equation}
and
\begin{equation}\label{jot-u-jeden-kwadrat-tylda}
\widetilde{J}[u^2_1]=\frac{N}{d^2}\left(\frac{
5d-3}{d+1}\right)\left(\frac{d-1}{d}\right)^{d-1},
\end{equation}
and in consequence
\begin{equation}
p_{\mathrm{POVM}}^{W,c}=\frac{3d-1}{d(d+1)}\left(\frac{d-1}{d}\right)^{d-1}.
\end{equation}
Interestingly, $p_{\mathrm{POVM}}^{W,c}>p^{W,c}_{\mathrm{sep}}$
for any $d$, and as a result Barrett's model reproduces statistics
of generalized measurements for entangled Werner states for any
$d$. However, unlike in Werner's model, $p_{\mathrm{POVM}}^{W,c}$
is a (monotonically) decreasing function of $d$. Moreover, it
decreases faster than $p^{W,c}_{\mathrm{sep}}$, meaning that the
range of $p$ for which this is the case, i.e.,
$p\in(p^{W,c}_{\mathrm{sep}},p^{W,c}_{\mathrm{POVM}}]$, shrinks
with $d\to\infty$. Fig. \ref{fig:Werner} compares the three
critical values $p^{W,c}_{\mathrm{sep}}$, $p^{W,c}_{\mathrm{PM}}$,
and $p^{W,c}_{\mathrm{POVM}}$ for various values of the local
dimension $d$.

\begin{rem}\label{remark_channels}
As noticed by Barrett \cite{Barrett}, a bipartite state $\rho\in
B(\mathbbm{C}^{d_A}\otimes\mathbbm{C}^{d_B})$ that has a local
model for generalized measurements induces a whole family of
states with local models (some of which might naturally be
separable). The construction goes as follows: let $\Omega$,
$\omega(\lambda)$, $p^1_A(a|\calA,\lambda)$ and
$p^1_B(b|\calB,\lambda)$ denote, respectively, the space of local
variables, the distribution and the response functions in the
local model for $\rho$ for general measurements
$\calA=\{\calA_a\}$ and $\calB=\{\calB_b\}$. Consider then two (in
general different) quantum channels\footnote{Recall that a linear
map $\Lambda:B(\mathcal{H})\to B(\mathcal{K})$ with $\mathcal{H}$
and $\mathcal{K}$ being two finite-dimensional Hilbert spaces is
called a \textit{quantum channel} iff it is completely positive
and trace-preserving. } $\Lambda_X:B(\mathbbm{C}^{d_X})\to
B(\mathbbm{C}^{d'_X})$, $X=A,B$, and the state obtained from
$\rho$ through
\beq \sigma=(\Lambda_A\ot\Lambda_B)(\rho). \eeq
Now, $\sigma$ has a local model for generalized measurements with the same
hidden state space $\Omega$, distribution $\omega$ and response
functions defined as
\begin{equation}\label{functions}
p^2_A(a|\calA,\lambda)=p^1_A(a|\calA',\lambda),\qquad
p^2_B(b|\calB,\lambda)=p^1_B(b|\calB',\lambda)
\end{equation}
with the measurements operators of the generalized measurements $\calA'$ and
$\calB'$
given by
\begin{equation}
\calA_a'=\Lambda_A^{\dagger}(\calA_a), \quad
\calB_b'=\Lambda_B^{\dagger}(\calB_b),
\end{equation}
where $\Lambda^{\dagger}$ is a dual\footnote{A dual map to a
linear map $\Lambda:B(\mathcal{H})\to B(\mathcal{K})$ is a linear
map $\Lambda^{\dagger}:B(\mathcal{K})\to B(\mathcal{H})$ that
satisfies $\Tr[X\Lambda(Y)]=\Tr[\Lambda^{\dagger}(X)Y]$ for all
$X\in B(\mathcal{K})$ and $Y\in B(\mathcal{H})$.}  map of
$\Lambda$. As a dual map of a quantum channel is positive and unital (it
preserves the identity operator), the operators $\calA_a'$ and $\calB_b'$
form proper quantum measurements. To see
eventually that the functions (\ref{functions}) do define a local
model for $\sigma$ it is enough to apply the following argument
\begin{eqnarray}
\int_{\Omega}\mathrm{d}\lambda\,\omega(\lambda)p^2_A(a|\calA,\lambda)
p^2_B(b|\calB,\lambda)&=&\Tr(\calA_a'\ot \calB_b'\rho)\nonumber\\
&=&\Tr[\Lambda_A^{\dagger}(\calA_a)\ot \Lambda_B^{\dagger}(\calB_b)\rho]\nonumber\\
&=&\Tr[\calA_a\ot \calB_b(\Lambda_A\ot\Lambda_B)(\rho)]\nonumber\\
&=&\Tr[\calA_a\ot \calB_b\sigma],
\end{eqnarray}
where to pass from the second to the third line we have exploited
the definition of the dual map to $\Lambda$.

Let us conclude by noting that the above argument can also be
applied to the multipartite states, provided such states with
local models for generalized measurements exist. Actually, it
applies to any state that has a ``mixed" local model, i.e., one
which works  for projective measurements at some sites and for
generalized measurements at the rest. In such case, to obtain a
new state local quantum channels discussed above can be applied
only to those sites. An example of such a state and the
corresponding ``mixed" local model will be discussed in Section
\ref{trzy-kubity-model}.
\end{rem}

\subsection{Nonlocality of noisy states: the Grothendieck
constant}\label{groth}

We now move to the analysis of nonlocal properties of noisy
quantum states, that is states of the form
\beq \label{noisy-state}\rho(d,p)=p\rho +(1-p)\frac{\jedynka
}{d^2},\quad \rho\in B(\cee{d}\otimes\cee{d}).\eeq
As mentioned at the beginning of this section, this problem can be
related to the mathematical constant $K_G$ known as  \grod
constant \cite{Grothendieck}. The connection between the latter
and nonlocality was first recognized in 1987 by Tsirelson (a.k.a.
Cirel'son) \cite{Tsirelson} who considered the problem of how
large is the set of quantum correlations compared to the set of
classical ones. The interest in this surprising relationship had
its revival almost thirty years later when it was analyzed in
greater detail by \acin \etal in Ref. \cite{Acin-grod}. The
results of the latter paper, relevant for our review, may be
summarized as follows (the terminology and necessary definitions
will be introduced in what follows):
\begin{itemize}
    \item[\numerek{1}] projective measurements on a two--qubit
    Werner state, $\rho_{\mathrm{W}}(2,p)$ (see Eq. (\ref{WernerState})), can be
simulated with an LHV model if
    and only if $p\le 1/K_G(3)$,
    \item[\numerek{2}] local models for $\rho_{\mathrm{W}}(2,p)$ exist in the
whole range
    where the state does not violate CHSH inequality (\ref{chsh-ineq}), that is for $p\le
    1/\sqrt{2}$, when at least one of the parties is restricted to perform
    planar measurements,
    \item[\numerek{3}]
whenever $p\le 1/K_G(2d^2)$ there is a local
    model reproducing joint correlations of traceless two-outcome observables for the
    state (\ref{noisy-state}) with {\it any} $\rho$. Moreover, for
$p>1/K_G(2\lfloor\log_2 d \rfloor+1)$ there exists $\rho(d,p)$
without such a model. In particular, in the limit of $d\to\infty$
both bounds match and every noisy state is local below $1/K_G$ and
this number cannot be made larger.
\end{itemize}

Before we proceed, we need the definition of the \grod constant.
Let $n\ge 2$ be integer, $M$ be an arbitrary $m\times m$ real
matrix such that for all real numbers $a_1,a_2,\cdots,a_m$,
$b_1,b_2,\cdots,b_m$ from the interval $[-1,+1]$, it holds
\begin{equation}
 \left|\sum_{i,j=1}^m M_{ij}a_i b_j\right|\le 1.
\end{equation}
The \grod constant of order $n$, $K_G(n)$, is defined to be the
smallest number such that
\beq\label{grodecka} \left|\sum_{i,j=1}^m
M_{ij}\bold{a}_i\cdot\bold{b}_j\right|\le K_G(n), \eeq
for all unit vectors $\bold{a}_k,\bold{b}_k\in \mathbb{R}^n$
 The \grod
constant, $K_G$, is then defined through
\beq K_G=\lim_{n\to\infty} K_G(n). \eeq
It is quite remarkable that such constant even exists; it is also
interesting that the exact values of the constants, besides
$K_G(2)=1/\sqrt{2}$ \cite{Krivine}, are not known. The bounds for
the constants appearing in our analysis are as follows
\cite{Fishburn,Krivine,Vertesi,Toner-doktorat}:
\beq\label{groth-bound} 1.6770 \le K_G \le 1.7822, \eeq
\beq\label{groth8-bound} K_G(8) \le 1.6641,     \eeq
\beq\label{groth3-bound} 1.417 \lesssim K_G(3)\le \frac{\pi}{2c_3},
\eeq
where $c_3$ is the unique, in the interval $[0,\pi/2]$, solution
of the equation
\begin{equation}\label{ce-trzy}
\sqrt{c_3}\int_{0}^{c_3} \mathrm{d}x\;x^{-3/2}\sin{x}=2.
\end{equation}
Numerically one then finds that
\beq\label{groth3-bound-numerical} K_G(3)\le 1.5163. \eeq
Let us now elaborate on each of the items \numerek{1}-\numerek{3}
from the list above.

As to the point \numerek{1}, let us first notice that the
reconstruction of mean values
$\langle\mathcal{A}\ot\mathcal{B}\rangle$,
$\langle\mathcal{A}\rangle$, and $\langle \mathcal{B}\rangle$ is
enough to retrieve full probability distribution as we deal here
with two-outcome measurements in which case the number of mean
values matches the number of independent probabilities necessary
for this task (see the discussion in Section \ref{preliminaries}).
Further, on states with maximally mixed reductions (for two qubits
these are the Bell diagonal states), including the discussed
Werner state $\rho_W(2,p)$, local expectation values vanish,
$\langle\mathcal{A}\rangle=\langle \mathcal{B}\rangle=0$. The key
point of the analysis now is that given a model reproducing
correctly only $\langle\mathcal{A}\ot\mathcal{B}\rangle$, we can
turn it into one for which also local mean values vanish: we
augment the old protocol with an extra random bit and in the new
model parties just multiply outputs of the old one by the value of
this bit ensuring that joint prediction are still the same and
local ones are vanishing. Thus, nonlocal properties of
$\rho_W(2,p)$ are uniquely determined by the joint correlations
solely and in the following analysis we can restrict ourselves to
{\it correlation Bell inequalities}. The latter in the most
general case can be written as
\begin{equation}\label{correlation_ineq}
\left|\sum_{i,j=1}^m M_{ij}\langle \calA_i\otimes
\calB_j\rangle\right|\le \beta_L,\quad M_{ij}\in\mathbb{R},
\end{equation}
where $\beta_L$ denotes the local bound of a given Bell inequality,
that is
\begin{equation}
 \beta_L=\max_{a_i,b_j=\pm 1} \left|\sum_{i,j=1}^m M_{ij}a_i b_j\right|
\end{equation}
since deterministic strategies are sufficient to achieve it.
Clearly, we can normalize our inequality such that $\beta_L=1$ and we
will assume this has been done.

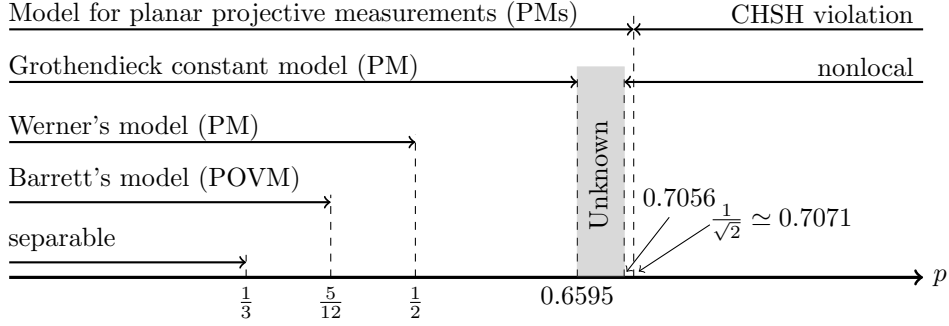
\begin{figure}[h!]

%%%%%%
\begin{tikzpicture}
%[xscale=15, yscale=10]
[xscale=13.5, yscale=10] \draw[black, very thick,->] (0.1,0) --
(1,0); \draw[black, very thick] (1,0) node[anchor=west] {$p$};
%linie przerywane
\draw[black,dashed] (1/3,0)-- (1/3,0.03);
\draw[black,dashed](5/12,0) -- (5/12,0.11);
 \draw[black,dashed] (1/2,0) -- (1/2,0.19);
\draw[black,dashed] (0.6595,0) -- (0.6595,0.27);
\draw[black,dashed] (0.7056,0) -- (0.7056,0.27);
\draw[black,dashed] (0.715,0) -- (0.715,0.35);
%strzalki
 \draw[black, <-, thick] (1/3,0.02)--(0.1,0.02);
\draw[black, <-, thick] (5/12,0.1)--(0.1,0.1);
 \draw[black, <-,thick] (1/2,0.18)--(0.1,0.18);
  \draw[black, <-,thick] (0.6595,0.26)--(0.1,0.26);
   \draw[black, <-,thick] (0.715,0.33)--(0.1,0.33);
    \draw[black, <-,thick] (0.715,0.33)--(1,0.33);
  \draw[black, <-,thick] (0.7056,0.26)--(1,0.26);
%strzalki do punktow na osi
 \draw[black, ->] (0.75,0.08)--(0.708,0.0075);
 \draw[black, ->] (0.79,0.07)--(0.72,0.0079);
%punkty na osi
\draw[black] (1/3,0) node[anchor = north] {$\frac{1}{3}$};
\draw[black] (5/12,0) node[anchor = north] {$\frac{5}{12}$};
\draw[black] (1/2,0) node[anchor = north] {$\frac{1}{2}$};
\draw[black] (0.6595,0) node[anchor = north] {$0.6595$};
%podpisy
\draw[black] (0.09,0.05) node[anchor = west] {separable};
\draw[black] (0.09,0.13) node[anchor = west] {Barrett's model
(POVM)}; \draw[black] (0.09,0.20) node[anchor = west] {Werner's
model (PM)}; \draw[black] (0.09,0.28) node[anchor = west]
{Grothendieck constant model (PM)}; \draw[black] (0.09,0.35)
node[anchor = west] {Model for planar projective measurements
(PMs)};\draw[black] (1,0.35) node[anchor = east] {CHSH violation};
\draw[black] (1,0.28) node[anchor = east] {nonlocal};
%podpisy punkty
\draw[black] (0.76,0.13) node[anchor = north] {$0.7056$};
\draw[black] (0.86,0.11) node[anchor = north]
{$\frac{1}{\sqrt{2}}\simeq 0.7071$};
%nieznany region
\draw [draw=gray,fill=gray,opacity=0.3] (0.6595,0.001) rectangle
(0.7056,0.28); \draw[black,thick] (0.68,0.05) node[anchor =
west,rotate = 90] {Unknown};
\end{tikzpicture}
%%%%%%
\caption{\textbf{Regions of the parameter $p$ in which the two-qubit Werner
states $\rho_W(2,p)$ are local/nonlocal}. Any state with $p\le 1/3$,
and only then, is separable and trivially has a local model.
Werner's model (Section \ref{werner-model}) for projective
measurements (PMs) works in the region $p\le 1/2$, Barrett's model
(Section \ref{baret}), which is designed for POVMs, is valid for
$p\le 5/12$. When one of the parties is restricted to perform
planar measurements there is a model (Section \ref{groth}) up to
$p=1/\sqrt{2}$, which is the threshold value above which
$\rho_W(2,p)$ violates the CHSH inequality. The model based on the
bounds on the \grod constant (Section \ref{groth}) works up to
$p=1/K_G(3)$, which with current state of the knowledge is not
smaller than $0.6595$. On the other hand, it is known that
$\rho_W(2,p)$ is nonlocal at least in the region $p\ge 0.7056$.
How large this gap is in reality is not known at this
moment.}\label{werner-regiony-modeli}
\end{figure}

Now, correlations on the maximally
mixed state vanish and thus violation of any correlation Bell
inequality by Werner states is determined by its violation by
the singlet state. Since for the latter we have $\langle
\calA_i\otimes \calB_j\rangle_{\psi_-} = \bold{a}_i \cdot
\bold{b}_j$, this violation maximized over all Bell inequalities
can be written just as
\begin{equation}
\beta_{Q,W}:=\lim_{m\to\infty}
\sup_{M_{ij}}\max_{\bold{a}_i,\bold{b}_j}\left|\sum_{i,j=1}^m
M_{ij}\bold{a}_i\cdot\bold{b}_j\right|.
\end{equation}
From this it immediately follows that no Werner state
$\rho_{W}(p,2)$ can violate a correlation Bell inequality in the
range $p\le 1/ \beta_{Q,W}$. As one can easily see,
$\beta_{Q,W}$ is just the \grod constant $K_G(3)$, and so we have
recovered that whenever $p\le 1/ K_G(3)$ a two-qubit Werner state
is local for projective measurements and no local model can exist
for values outside this range. Taking into account
(\ref{groth3-bound-numerical}) we realize that this result
is an improvement over Werner's $p\le 1/2$, as $1/K_G(3)\geq
0.6595$. It is known that there is a gap between the exact value
and $1/\sqrt{2}\simeq 0.7071$ as the Werner state have been
demonstrated to violate some Bell inequality for $p> 0.7056$
\cite{Vertesi} (cf. Fig. \ref{werner-regiony-modeli}). Recall that
$\rho_W(2,p)$ is entangled when $p>1/3$.

As announced in \numerek{2}, there is no such separation when (at
least) one of the parties is restricted to perform measurements on
a plane in the Bloch sphere. In this case vectors $\bold{a}_i$ are
two dimensional, moreover, $\bold{b}_j$ can be taken to lie in the
same plane since only the scalar products of $\bold{a}$ and
$\bold{b}$ are contributing to the value of the Bell operator. In
this way it is clear that the bound now involves $K_G(2)$, which,
as mentioned, is equal to $1/\sqrt{2}$. The result, which is an
improvement over $2/\pi$ from \cite{Larsson}, then follows.

The statement of \numerek{3} is a result of the attempt to
generalize \numerek{1} to arbitrary dimensions. However, we
encounter two difficulties in such generalization. First, we need
to restrict ourselves to two-outcomes measurements (for the reason
discussed in Section \ref{preliminaries}). Second, even in this
case we cannot hope to give full characterization of nonlocal
properties of a state with the aid of the correlation Bell
inequalities only. This stems from the fact that in the general
approach we pursue here we cannot assume that our states are
locally maximally mixed and thus it might be necessary to use
inequalities with local terms \cite{CollinsGisin} for this
purpose. Thus, we can only hope here to fully characterize the
joint correlations. Further, an additional requirement will have
to be met: tracelessness of the observables. This condition
ensures that the averages on the maximally mixed state are zero
and the \grod constant approach can be employed.

To proceed it might be useful to rephrase \numerek{3} using the
critical probability $p^c(d)$ for states of the form Eq.
(\ref{noisy-state}). For a given $d$, this is a minimum over all
states $\rho$ of the maximal $p$ for which there exists a local
model in the given scenario. With this notion in hand, the claims
are that
\begin{equation}\label{boundy-groth}
\frac{1}{K_G(2d^2)}\le p^c(d) \le \frac{1}{K_G(2\lfloor\log_2 d
\rfloor+1)}
\end{equation}
and
\beq p^c(\infty):=\lim_{d\to\infty}p^c(d)=\frac{1}{K_G}. \eeq

We start with the left-hand side of (\ref{boundy-groth}) and
consider first states $\rho^{(i)}(d,p)$ of the form (\ref{noisy-state}) with
$\rho$
being an arbitrary pure state $\ket{\psi_i}$. From Refs.
\cite{Tsirelson,Acin-grod} we know that for any observables
$\calA$, $\calB$ and a state $\ket{\psi}\in
\cee{d}\otimes\cee{d}$, one can find vectors $\bold{a}$ and
$\bold{b}$ from $\mathbb{R}^{2d^2}$ such that
\beq
 \langle \calA\otimes \calB  \rangle =\bold{a}\cdot \bold{b}.
\eeq
  With the definition of the \grod constant in mind, we conclude
that the critical probability for states $\rho^{(i)}(d,p)$ is at
least equal to $1/K_G(2d^2)$. Since any $\rho$ can be expressed as
a convex combination of pure states, $\rho=\sum_i q_i
\proj{\psi_i}$, the bound we have just established for
$\rho^{(i)}(d,p)$ must also hold for general $\rho_p$. This
follows from the fact that our local model for these states might
be just taken to be the convex combination of models for
$\rho^{(i)}(d,p)$ with weights $q_i$. In conclusion, whenever $p
\le 1/K_G(2d^2)$ the state $\rho(d,p)$ is local regardless of the
form of $\rho$.

Now let us move to the right-hand side of (\ref{boundy-groth}).
Assume we have a set of unit vectors $\bold{a}_i,\bold{b}_j\in\mathbbm{R}^n$
with $1\le i,j \le m$ and $n=2 \lfloor \log d\rfloor+1$, maximizing the
left-hand side of Eq. (\ref{grodecka}), that is achieving $K_G(2 \lfloor \log
d\rfloor+1)$. It is known that there exist traceless observables
$\calA_i$, $\calB_j$ such that $\langle \calA_i \otimes \calB_j
\rangle_{\phi_+} = \bold{a}_i \cdot \bold{b}_j$ for
the two-qudit maximally entangled state
\begin{equation}\label{maksymalnie}
 \ket{\phi_{+}^d}=\frac{1}{\sqrt{d}}\sum_{i=0}^{d-1}\ket{ii}
\end{equation}
with
$d=2^{\lfloor n/2 \rfloor}$. In Eq. (\ref{noisy-state}), take now
$\rho=\proj{\phi_{+}^d}$ and $p=(1+\epsilon)/K_G(2\lfloor \log d
\rfloor +1)$ for some $\epsilon
>0$. From the above it follows that we will
achieve $1+\epsilon$ for the value of the Bell operator which
means the violation of a Bell inequality and rules out existence
of a local model for these states. To sum up, for $p>
1/K_G(2\lfloor \log d \rfloor +1)$ there exist states
(\ref{noisy-state}) with nonlocal joint correlations.

Taking the limit of both sides clearly results in the
threshold value equal to $1/K_G$.
This concludes the proof of the claims made in \numerek{3}.

Let us now comment about possible applications of \numerek{3}. We
list them below and then explain the underlying reasoning.
\begin{itemize}
\item[\numerek{3.1}] for the noisy qubit states $\rho(2,p)$
of the form (\ref{noisy-state}) there always exists a
local model for joint correlations whenever $p\le 1/K_G(8)$,
\item[\numerek{3.2}] for the isotropic states (that is, those with
$\rho=\proj{\phi_+^d}$; see also Sec. \ref{almeida}) there is a local
model simulating the full probability distribution for traceless
observables whenever $p\le 1/K_G(d^2-1)$.
\end{itemize}

The statement of \numerek{3.1} follows from the fact that in the
qubit scenario observables are two-outcome and traceless, just as
required. Taking into account (\ref{groth8-bound}) this gives the
threshold around $0.6009$. On the other hand, \numerek{3.2} stems
from the following: (i)  previously mentioned possibility of
adding extra random bit to the protocol  to reproduce local mean
values, which are zero for the considered state, and (ii) a
refinement of one of the facts mentioned earlier, namely, when the
state is maximally entangled, to reproduce mean values in the form
$\bold{a}\cdot \bold{b}$ both vectors can be drawn from
$\mathbb{R}^{d^2-1}$.

Interestingly, the results by Grothendieck
\cite{Grothendieck} and Krivine \cite{Krivine}, or more precisely their
proofs concerning upper bounds on the \grod constant, allowed Toner
\cite{Toner-doktorat} to obtain explicit local models for the
above considered cases. We conclude this section by giving one of these models
with a sketch of the proof (details can be found in \cite{Toner-doktorat}),
namely the one for projective measurements on the two--qubit Werner states in
the range $p<0.6595$.

Let $\bold{a},\bold{b}$ be unit vectors from $\mathbb{R}^{3}$
representing Alice and Bob measurements.
Let further $f$ and $g$ be mappings $\mathbb{R}^{3}\to
\bigoplus_{k=0}^{\infty}\bigoplus_{m=-(2k+1)}^{2k+1}
\mathbb{R}^2$ defined by the following set of equations:
\begin{eqnarray}\label{toner-Alicja}
f(\boldsymbol{\mathrm{a}}) &=&
\bigoplus_{k=0}^{\infty}\bigoplus_{m=-(2k+1)}^{2k+1} f_{2k+1,m}
(\boldsymbol{\mathrm{a}}),\nonumber \\
f_{2k+1,m} (\boldsymbol{\mathrm{a}})  &=& (-)^{k+1}
\sqrt{\frac{4\pi^{3/2}J_{2k+3/2}(c_3)}{\sqrt{2c_3}}}
\left[ \mathrm{Re}( Y_{2k+1}^m (\boldsymbol{\mathrm{a}})), \mathrm{Im}
(Y_{2k+1}^m
(\boldsymbol{\mathrm{a}})) \right]\non
\end{eqnarray}
and
\begin{eqnarray}\label{toner-Bob}
g(\boldsymbol{\mathrm{b}}) &=&
\bigoplus_{k=0}^{\infty}\bigoplus_{m=-(2k+1)}^{2k+1} g_{2k+1,m}
(\boldsymbol{\mathrm{b}}),\nonumber \\
g_{2k+1,m} (\boldsymbol{\mathrm{b}})  &=&
\sqrt{\frac{4\pi^{3/2}J_{2k+3/2}(c_3)}{\sqrt{2c_3}}} \left[
\mathrm{Re} (Y_{2k+1}^m (\boldsymbol{\mathrm{b}})), \mathrm{Im}
(Y_{2k+1}^m (\boldsymbol{\mathrm{b}})) \right],
\end{eqnarray}
where $c_3$ is defined through (\ref{ce-trzy}),
$Y_{l}^{m}$ are the spherical harmonics, and $J_{\nu}$ are the
Bessel functions of the first kind of order $\nu$. The model reads
as follows \footnote{Notice that the model fits the general
scheme given in Section \ref{preliminaries} since we can write the
corresponding response functions as $\label{ProbA} p(\pm
1|\calA,\lambda)=\left\{
\begin{array}{ll}
1, \quad  \mathrm{sgn} (f(\boldsymbol{\mathrm{a}})\cdot
 \boldsymbol{\lambda})=\pm 1 \\
0,\quad \mathrm{sgn} (f(\boldsymbol{\mathrm{a}})\cdot
 \boldsymbol{\lambda})=\mp 1
\end{array}
\right.$ and analogously for Bob.}:

\linka
\begin{center}
\textbf{LHV model for projective measurements for the two--qubit
Werner states \cite{Toner-doktorat}}
\end{center}
\begin{itemize}
\item[(1)] Alice and Bob each get an infinite sequence of numbers
$\boldsymbol{\lambda}=[\lambda_1,\lambda_2,\ldots]\in\mathbbm{R}^{\infty}$,
where each $\lambda_i$ is drawn from a normal
distribution with the mean equal to $0$ and the standard deviation
equal to $1$,
\item[(2)] Alice outputs  $a=\mathrm{sgn} (f(\boldsymbol{\mathrm{a}})\cdot
 \boldsymbol{\lambda})$ with $f$ defined through Eq.
 (\ref{toner-Alicja}),
 \item[(3)]   Bob outputs $b=\mathrm{sgn}( g(\boldsymbol{\mathrm{b}})\cdot
 \boldsymbol{\lambda})$ with $g$ defined through Eq.
 (\ref{toner-Bob}).
\end{itemize}
\linka

Let us now recall some important steps from the proof
\cite{Toner-doktorat} that the model works in the required range.

First one needs to verify that $f(\boldsymbol{\mathrm{a}})$ is a unit
vector (for $g(\boldsymbol{\mathrm{b}}$) the reasoning will be similar so
in further parts we omit it) . One obtains:
\beqn f(\boldsymbol{\mathrm{a}}) \cdot
f(\boldsymbol{\mathrm{a}})&=&\sqrt{\frac{\pi}{2c_3}}
\sum_{k=0}^{\infty}(4k+3) J_{2k+3/2}(c_3) \non
 &=& \frac{\sqrt{c_3}}{2} \int_{0}^{c_3}
\mathrm{d}x\;x^{-3/2}\sin{x}\non
&=& 1\eeqn
with the last equality following from the definition of $c_3$.
Further:
\beq f(\boldsymbol{\mathrm{a}})\cdot g(\boldsymbol{\mathrm{\mathrm{b}}})=
\sqrt{\frac{\pi}{2c_3}} \sum_{k=0}^{\infty}(-)^{k+1}(4k+3)
J_{2k+3/2}(c_3) P_{2k+1} (\boldsymbol{\mathrm{a}} \cdot \boldsymbol{\mathrm{b}}
),
\eeq
where $P_{l}$ are the Legendre polynomials. One further verifies
that
\beq -\sin (c_3 x)=  \sum_{k=0}^{\infty}(-)^{k+1}(4k+3)
J_{2k+3/2}(c_3) P_{2k+1} (x)\eeq
which in turn means that
\beq  f(\boldsymbol{\mathrm{a}})\cdot g(\boldsymbol{\mathrm{b}})= -\sin (c_3\,
\boldsymbol{\mathrm{a}}\cdot \boldsymbol{\mathrm{b}}). \eeq
Now, given the random variable $\boldsymbol{\lambda}$ specified by
the conditions given in the frame above and $a=\mathrm{sgn}
(\bold{x}\cdot\boldsymbol{\lambda})$, $b=\mathrm{sgn}
(\bold{y}\cdot\boldsymbol{\lambda})$, the average of $ab$ over
$\boldsymbol{\lambda}$  is equal $\langle ab \rangle =
(2/\pi)\sin^{-1}(\bold{x}\cdot \bold{y})$ for arbitrary
$\boldsymbol{\mathrm{x}},\boldsymbol{\mathrm{y}}\in
\mathbb{R}^{\infty}$ \cite{Grothendieck,Bell,Toner-doktorat}. We
thus find that:
\beq \langle \mathcal{A}\ot\mathcal{B}\rangle = - \frac{2c_3}{\pi}
\boldsymbol{\mathrm{a}}\cdot \boldsymbol{\mathrm{b}}.  \eeq
Having in mind (\ref{groth3-bound}) and
(\ref{groth3-bound-numerical})  we conclude correctness of the
model in the claimed range $p<0.6595$.

The results of this and previous sections
concerning the existence of local models for two-qubit Werner
states are summarized in Fig. \ref{werner-regiony-modeli}.

Let us conclude by noting that with the aid of the
\textit{complex} Grothendieck constant it is possible to
generalize some of the above statements to Bell inequalities with
an arbitrary number of outcomes \cite{JRA}.

\subsection{Noise robustness of correlations: Almeida \etal's
model}\label{almeida} The research on the robustness of
nonlocality in a general scenario was further pursued by Almeida
\etal \cite{Almeida}. The starting point of their
analysis was to check to what extent the nonlocality of maximally
entangled states of arbitrary dimension is affected by white
noise. That is, the goal was to determine when a local model
exists for states of the form
\begin{equation}\label{isotropic}
\rho_{\mathrm{iso}}(d,p)=p\proj{\phi_+^d}+(1-p)\frac{\mathbbm{1}_{d^2}}{d^2},
\end{equation}
where $0\leq p\leq 1$ and $\ket{\phi_+^d}$ is a maximally
entangled state (\ref{maksymalnie}).
These states are known in the literature as isotropic states and
are the unique states which are invariant under bilateral unitary
rotations of the form $U\otimes U^{\star}$. Note that isotropic
states have a clear physical meaning, as they correspond to noisy
versions of maximally entangled states, while this physical
interpretation is missing for Werner states of dimension larger
than two. In the case of qubits, isotropic and Werner states are
equivalent up to local unitary transformations.

Inspired directly by Werner's construction (see Section
\ref{werner-model}) the local model simulating projective
measurements $\calA$ and $\calB$ with measurement operators $\{
P_a \}$ and $\{ Q_b \}$, respectively, on the isotropic states is
given by:

\linka
\begin{center}
\textbf{LHV model for projective measurements on the isotropic
states \cite{Almeida,WisemanSteering}}
\end{center}
\begin{itemize}
\item[(1)] Alice and Bob each get $\ket{\lambda}\in
\Omega_d=\{\ket{\lambda}\in\mathbbm{C}^d|\langle\lambda|\lambda\rangle=1\}$
with the uniform distribution $\omega_d$,
\item[(2)] Alice's response function is:
\begin{equation}\label{almdeidaA}
p_{\mathrm{iso}}(a|\calA,\lambda)=\left\{
\begin{array}{ll}
1, \quad \mathrm{if}\;
\langle\lambda|P_a|\lambda\rangle=\max_\alpha\langle\lambda|P_{\alpha}
|\lambda\rangle\\
0,\quad\mathrm{otherwise}
\end{array},
\right.
\end{equation}
 \item[(3)] Bob's response function is:\begin{equation}\label{almeidaB}
p_{\mathrm{iso}}(b|\calB,\lambda)=\langle\lambda|Q^T_b|\lambda\rangle,
\end{equation}
where $T$ stands for the transposition.
\end{itemize}
\linka

\begin{figure*}[h!]
\centering\includegraphics[clip,width=0.5\columnwidth]{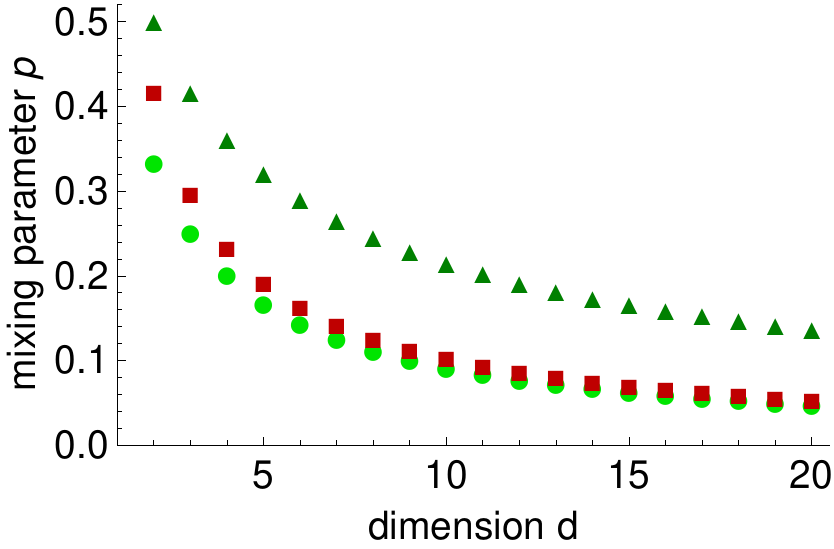}
\caption{\textbf{Critical probabilities for the isotropic state.}
Comparison of three critical values of probabilities for the
isotropic states (\ref{isotropic}) for $2\leq d\leq 20$:
$p^{\mathrm{iso,c}}_{\mathrm{sep}}$ (green dots),
$p^{\mathrm{iso,c}}_{\mathrm{POVM}}$ (red squares),
$p^{\mathrm{iso,c}}_{\mathrm{PM}}$ (green triangles). Notice that
contrary to the case of the Werner states, here the critical probability
$p^{\mathrm{iso,c}}_{\mathrm{PM}}$ drops with
$d$.}\label{fig:Iso}
\end{figure*}

Direct calculation, which can be carried out with the techniques
analogous to those already presented in Section
\ref{werner-model}, shows
that the probabilities realized by this model assume the same form
(\ref{Castelldefels2}) with $Q_b$ replaced by $Q_b^T$ and the
integral $J[h]$ replaced by $\mathcal{J}[h]$ given by (notice the
change of the integration ranges)
\begin{eqnarray}
\mathcal{J}[h]=\int_{0}^1\mathrm{d}u_1\int_{0}^{u_1}\mathrm{d}u_2\ldots
\int_{0}^{u_1}\mathrm{d}u_d
h(u_1,\ldots,u_d)\delta(u_1+\ldots+u_d-1).\nonumber\\
\end{eqnarray}
As before, $\mathcal{J}[1]$ can be directly determined from
(\ref{Castelldefels2}) and amounts to $N/d$, while
$\mathcal{J}[u_1]$ is computed in \ref{appendix} and is
proportional to the so-called harmonic number, i.e.,
\begin{equation}\label{jot-cal-u-jeden}
 \mathcal{J}[u_1]=\frac{N}{d^2}\sum_{k=1}^d\frac{1}{k}.
\end{equation}
After inserting all this into (\ref{Castelldefels2}), one arrives
at the following critical probability
\beq p=p^{\mathrm{iso},c}_{\mathrm{PM}}\equiv\frac{1}{d-1}\left(
-1 + \sum_{k=1}^d\frac{1}{k} \right) \eeq
for which the above model reproduces the statistics of the
isotropic states, meaning that the isotropic state is local at
least up to $p^{\mathrm{iso},c}_{\mathrm{PM}}$. Clearly, for $d=2$
this reproduces the critical value $1/2$ by Werner as it should
since then, as noted above, the isotropic and the Werner states
are related to each other via local unitary rotations. Noticeably,
in the limit of large $d$, the above critical probability scales
as $\log d/ d$. On the other hand it is known
\cite{HorodeckiReduction} that isotropic states are separable
whenever $p\le p_{\mathrm{sep}}^{\mathrm{iso,c}}$ with
\beq p_{\mathrm{sep}}^{\mathrm{iso,c}}\equiv \frac{1}{d+1} \eeq
and thus the critical probability for the local model is asymptotically $\log d$
larger than the corresponding value for entanglement (cf. Fig. \ref{fig:Iso}).

We can now move to the general case of arbitrary $\rho$. As it was
already noted in Section \ref{groth} it is enough to construct a
model for a pure state $\rho=\proj{\psi}$ as the model for a mixed
state can be taken to be a convex combination of models for pure
states.

The key point of the general approach is the local model for the
following mixture of $\rho$ with a state-dependent noise
\beq\label{state dependent noise}
\tilde{\varrho}=p^{\mathrm{iso},c}_{\mathrm{PM}}\proj{\psi}
+(1-p^{\mathrm{iso},c}_{\mathrm{PM}})\sigma\otimes\frac{\jedynka
}{d},\eeq
where $\sigma=\tr _B \proj{\psi}$. This state can be further
transformed into a one of the form (\ref{noisy-state}) by admixing
it with the following separable state
\begin{equation}
 \frac{1}{d-1}\sum_{k=1}^{d-1}\sigma_k\ot\frac{\mathbbm{1}}{d},
\end{equation}
where $\sigma_k=\sum_j \alpha_{j+k(\mathrm{mod}d)}\proj{j}$ with
$\{\alpha_j,\ket{j}\}$ being the eigensystem of $\sigma$.
The resulting state then reads
\beq \Theta= q \tilde{\varrho}+
\frac{1-q}{d-1}\sum_{k=1}^{d-1}\sigma_k \otimes
\frac{\jedynka}{d}.\eeq
It is easy to see that this state is of the desired form
(\ref{noisy-state}) when the weights fulfill the condition
$q(1-p)=(1-q)/(d-1)$. Since the state $\tilde{\varrho}$ has been
shown to be local, the state $\Theta$ is also local as it is just
a convex combination of local states. The condition on $q$,
however, implies that the price we have to pay for this
transformation is the diminishment of the value of the critical
probability, denoted by $\tilde{p}^{c}_{\mathrm{PM}}$, for which
we can construct a model.

Let us now move to the details of the construction of the model
for $\tilde{\varrho}$. The main tool is Nielsen's protocol
for the LOCC conversion between pure states \cite{Nielsen}. The idea is
that some preprocessing based on such transformation can be
performed by the source itself on the hidden state $\ket{\lambda}$
producing with some probabilities $\ket{\lambda^A_i}$ and
$\ket{\lambda^B_i}$, which are later sent to the parties. The
parties then follow the protocol for the isotropic state (see the
frame above) taking these hidden states instead of the standard
one. Thus, having in hands the model for the isotropic states we
get almost for free the model for the noisy states
$\tilde{\varrho}$.

To understand the details we need to see how
the conversion of $\ket{\phi_+^d}$ to some $\ket{\psi}$ works.
Assume that in the Schmidt form the state $\psi$ reads
$\ket{\psi}=\sum_k s_k \ket{k}\ket{k}$. Denoting $S=\mathrm{diag}
(s_0,s_1, \cdots,s_{d-1})$ and $U_k=\sum_{j=0}^{d-1}
\ket{j}\bra{j+k(\mathrm{mod} d)}$ with $k=0,1,\cdots,d-1$,
we can write
\beq \ket{\psi}=\sqrt{d} (X_k \otimes U_k)\ket{\phi_+^d},\quad
X_k=S\: U_k.\eeq Observe that $M_k\equiv X_k^{\dagger}X_k \geq 0$
and $\sum_k M_k =\jedynka_d$, which means that the operators $M_k$
constitute valid elements of a POVM measurement; let us call this measurement
$\calM$. The conversion protocol appears obvious now: Alice
performs $\calM$ with probability $\bra{\phi_+^d}M_k\ket{\phi_+^d}
=1/d$ obtaining the outcome $k$, she then sends the index of the
obtained result to Bob who performs appropriate unitary rotation
$U_k$, which in turn results in sharing $\ket{\psi}$ between the
parties.

As announced earlier, some preprocessing is simulated at the
source before the distribution stage of the protocol. More
precisely, the measurement $\calN$ with elements $N_k\equiv
X_k^{T}X_k^*$ on the hidden state $\ket{\lambda}$ is simulated.
The outcomes are obtained with probabilities
$p_{k,\lambda}=\bra{\lambda}N_k \ket{\lambda} $ and with these
probabilities the following states (more precisely their classical
descriptions) are sent to Alice and Bob: $\ket{\lambda^A_k}\equiv
(X_k^*/\sqrt{p_{k,\lambda}}) \ket{\lambda}$ and
$\ket{\lambda^B_k}\equiv U_k \ket{\lambda}$. Alice and Bob give
their outputs according to the response functions
$\tilde{p}(a|\calA,\lambda^A_i)$ and
$\tilde{p}(b|\calB,\lambda^B_i)$, respectively, which
are the same as in the protocol the isotropic states but with
$\lambda_A$ and $\lambda_B$ replacing $\lambda$. The statistics
they generate are now ($\Omega_d$ and $\omega_d$ are the
same as in Werner's model):
\beq \widetilde{p}_L(a,b|\calA,\calB)= \int_{\Omega_d}
\mathrm{d}\lambda\; \omega_d(\lambda) \sum_{i=0}^{d-1}
p_{i,\lambda} \widetilde{p}(a|\calA,\lambda^A_i)
\widetilde{p}(b|\calB,\lambda^B_i), \eeq
 and one can easily show that they indeed reproduce
quantum prediction for $\tilde{\varrho}$.

To complete the protocol for the general state one must add noise
to Alice's share which can be done as mentioned above. The model
for arbitrary noisy state is summarized in the frame below (see
the text above for the notation):

\linka
\begin{center}
\textbf{LHV model for projective measurements on general
noisy state (\ref{noisy-state}) \cite{Almeida} }
\end{center}
The parties follow the protocols $\calP$ and $\calQ$ listed below
with respective fractions $q$ and $1-q$ of times.
\begin{center}
Protocol $\calP$:
\end{center}
\begin{itemize}
\item[($\calP_0$)] $\ket{\lambda}$ are drawn from
$\Omega_d$ according to the uniform distribution $\omega_d$,
\item[($\calP_1$)] with the probability  $p_{k,\lambda}=\bra{\lambda}N_k \ket{\lambda} $ Alice gets $\ket{\lambda^A_k}\equiv
(X_k^*/\sqrt{p_{k,\lambda}}) \ket{\lambda}$ and Bob gets
$\ket{\lambda^B_k}\equiv U_k \ket{\lambda}$,
\item[($\calP_2$)] Alice's response function is \begin{equation}\label{ProbAnoisy}
\widetilde{p}(a|\calA,\lambda_k^A)=\left\{
\begin{array}{ll}
1, \quad \mathrm{if}\,\,
\langle\lambda_k^A|P_a|\lambda_k^A\rangle=\max
_\alpha\langle\lambda_k^A|P_{\alpha}
|\lambda_k^A\rangle\\
0,\quad\mathrm{otherwise}
\end{array},
\right.
\end{equation}
 \item[($\calP_3$)] Bob's response is:\begin{equation}\label{ProbBnoisy}
\widetilde{p}(b|\calB,\lambda_k^B)=\langle\lambda_k^B|Q^T_b|\lambda_k^B\rangle.
\end{equation}
\end{itemize}
\begin{center}
Protocol $\calQ$:
\end{center}
\begin{itemize}
\item[($\calQ_1$)] Alice simulates the statistics of measurements on $\sum_{k=1}^{d-1}\sigma_k/(d-1)$,
\item[($\calQ_2$)] Bob outputs random results simulating in this way the statistics of measurements on the maximally mixed
states.
\end{itemize}
\linka

As noted above the threshold value $\tilde{p}^c_{\mathrm{PM}}$ for the model
presented above is now reduced in comparison to the one for the
isotropic state and is equal
\beq\tilde{p}^c_{\mathrm{PM}}=\frac{p^{\mathrm{iso},c}_{\mathrm{PM}}}{(1-p^{\mathrm{iso},c}_{\mathrm{PM}})(d-1)+1}.
\eeq
In the limit of large $d$ this scales as $\log d/d^2$. In
comparison, the threshold value $\tilde{p}_{\mathrm{sep}}^c$ for
the separability of states (\ref{noisy-state}) lies in the
following interval \cite{Gurvits}:
\beq \tilde{p}_{\mathrm{sep}}^c\in  \left[
\frac{1}{d^2-1},\frac{2}{d^2+2}\right]. \eeq
This means, as previously, that the locality threshold is at least
asymptotically $\log d$ larger than the corresponding separability value.

Exploiting Barrett's model (see Section \ref{baret}), the ideas
presented above can be adapted to the general case of
POVMs. The resulting protocol for the general noisy states is as follows:

\linka
\begin{center}
\textbf{LHV model for POVMs on the  noisy states
(\ref{noisy-state}) \cite{Almeida} }
\end{center}
Fraction $q$ of times the parties follow the below described
protocol $\calP'$ and fraction $1-q$ the admixing protocol $\calQ$
described in the frame above.
\begin{center}
Protocol $\calP'$:
\end{center}
\begin{itemize}
\item[($\calP'_0$)] $\ket{\lambda}$ are drawn from
$\Omega_d$
according to the distribution $\omega_d(\lambda)$,
\item[($\calP'_1$)] with the probability  $p_{k,\lambda}=\bra{\lambda}N_k \ket{\lambda} $ Alice gets $\ket{\lambda^A_k}\equiv
(X_k^*/\sqrt{p_{k,\lambda}}) \ket{\lambda}$ and Bob gets
$\ket{\lambda^B_k}\equiv U_k \ket{\lambda}$,
\item[($\calP'_2$)] Alice's response function is
\begin{eqnarray}\label{ProbAgeneral}
\hspace{-1cm} p'(a|\calA,
\lambda_k^A)&=&\langle\lambda_k^A|\calA_a|\lambda_k^A\rangle\Theta(\langle\lambda_k^
A|P_a|\lambda_k^A\rangle-\textstyle{\frac{1}{d}})\nonumber
\\&&+\left[
1-\sum_j\langle\lambda_k^A|\calA_j|\lambda_k^A\rangle\Theta(\langle\lambda_k^A|P_j|
\lambda_k^A\rangle-\textstyle{\frac{1}{d}})\right]\frac{\eta_a}{d}, \nonumber\\
\end{eqnarray}
 \item[($\calP'_3$)] Bob's response is:\begin{equation}\label{ProbBgeneral}
p'(b|\calB,\lambda_k^B)=\xi_b\langle\lambda_k^B|Q^T_b|\lambda_k^B\rangle.
\end{equation}
\end{itemize}
\linka

The steps $(\calP'_2)$-$(\calP'_3)$ are the ones required to
simulate statistics for the isotropic state and obviously could be
performed separately on $\ket{\lambda}$ if this was the task. The
step $(\calP'_1)$, as previously, originates from Nielsen's
protocol. We need to combine  the protocol with $\calQ$ to ensure
that the resulting state has the proper form.

Let us now take a closer look at the range of parameters for which
the model works. The critical value
$p_{\mathrm{POVM}}^{\mathrm{iso,c}}$  for the isotropic state is
\beq
p_{\mathrm{POVM}}^{\mathrm{iso,c}}=\frac{(3d-1)(d-1)^{d-1}}{(d+1)d^d},
\eeq
which for large $d$ scales as $3/(\mathrm{e}d)$. In the general
case (i.e., arbitrary $\rho$ in (\ref{noisy-state})) the critical
threshold value, in analogy to the projective measurement case, is
\beq
\tilde{p}_{\mathrm{POVM}}^{c}=\frac{p_{\mathrm{POVM}}^{\mathrm{iso,c}}}{(1-p_{\mathrm{POVM}}^{\mathrm{iso,c}}
)(d-1)+1} ,\eeq
which asymptotically scales as $3/(\mathrm{e}d^2)$.

\subsection{From projective to
generalized measurements --- Hirsch \etal's
construction}\label{pm-to-povm}

An interesting approach to the construction of states with local
models for POVMs was put forward by Hirsch \etal \cite{Hirsch}.
The innovation of their construction consisted in the somewhat
different logic compared to the constructions reported
above. As already discussed in Section \ref{baret}, it had been
known that by applying a local channel to a state with a local
model for POVMs one obtains another state with an underlying local
model for the same type of measurements. Hirsch \etal proposed a
method of constructing states with local models for {\it
arbitrary} measurements departing from states with models for {\it
projective} measurements.

Assume $\varrho_0\in B(\cee{d}\otimes\cee{d})$ has a local model
for any dichotomic projective measurements
with measurement operators given by $\{P_a,\jedynka-P_a\}$ and
$\{Q_b,\jedynka-Q_b\}$ for Alice and Bob, respectively.
In what follows we will show that the state
\beq \label{stan_z_POVM}\varrho=\frac{1}{d^2}\Big\{
\varrho_0+(d-1)
(\varrho_A\otimes\sigma_B+\sigma_A\otimes\varrho_B)+
(d-1)^2\varrho_A\otimes\sigma_B \Big\},
 \eeq
where $\varrho_{A,B}$ are reductions of the original state
$\varrho_0$ and $\sigma_{A,B}$ are arbitrary, is local for
arbitrary POVMs with elements (see Section \ref{preliminaries})
$\calA_a=\eta_a P_a$ and $\calB_b=\xi_b Q_b$ respectively for Alice and
Bob. The proof of this fact relies on the explicit construction of
the corresponding local model, which is the following: \linka
\begin{center}
\textbf{LHV model for POVMs for $\varrho$ from Eq.
(\ref{stan_z_POVM}) \cite{Hirsch}}
\end{center}
\begin{itemize}
\item[(1)] Alice (Bob) chooses $P_a$ ($Q_b$) with the probability $\eta_a/d$
($\xi_b/d$),
\item[(2)] they simulate the measurement of the dichotomic observables
$\widetilde{\calA}_a=P_a- P_a^{\perp}$ and
 $\widetilde{\calB}_b=Q_b-Q_b^{\perp}$, respectively, on the state $\varrho_0$
with $P_a^{\perp}=\mathbbm{1}-P_a$ and $Q_b^{\perp}=\mathbbm{1}-Q_b$,
 \item[(3)] if the result in step (2) is $+1$, Alice (Bob) announces $a$ ($b$)
as the result of the simulation of the measurement on $\varrho$,
\item[(4)] if the result in step (2) is $-1$,  Alice (Bob) gives
arbitrary $a$ ($b$) as the output with the probability
$\tr(\sigma_A \calA_a)$ ($\tr(\sigma_B \calB_b)$).
\end{itemize}
\linka

 Let us
now argue that indeed this model correctly reproduces the quantum
probability, which, as can be easily verified, reads
\beqn\label{hirsz}
\fl p_Q(a,b|\calA,\calB)= \frac{\eta_a
\xi_b}{d^2}\Big\{ \tr[(P_a\otimes Q_b)\varrho_0]+
 (d-1)^2 \tr(P_a\sigma_A) \tr( Q_b \sigma_B) \non
 \hspace{1cm}+(d-1)\left[\tr(
P_a\varrho_A) \tr( Q_b \sigma_B) +\tr(P_a\sigma_A) \tr( Q_b \rho_B)
\right]
 \Big\}
 \eeqn
Both Alice and Bob, following the above protocol, may possibly
output either in step (3) or (4), resulting in four probabilities
of outputting the pair of outcomes $(a,b)$.

Let us begin with the case of both Alice and Bob outputting in
step (3). Since the simulation in this step concerns the original
state $\varrho_0$, this will happen with the probability $(\eta_a
\xi_b/d^2)\tr[(P_a\otimes Q_b)\varrho_0]$, which is the first term
in (\ref{hirsz}). Other possibility is that Alice produces an
output in step (3) but Bob fails to do the same and outputs in the
next step. As one can easily verify this will occur with the
probability $[(d-1)/d^2] \tr(\mathcal{A}_a\varrho_A) \tr(\calB_b
\sigma_B)$. On the other hand, the event when Alice produces the
output in the last step but Bob does it in the third one, will
occur with the probability $[(d-1)/d^2] \tr(\calA_a\sigma_A)
\tr(\calB_b\varrho_B)$; these two probabilities are, respectively,
the third and the fourth term in Eq. (\ref{hirsz}). The remaining
case of both Alice and Bob outputting in the last step will happen
with the probability $[(d-1)/d^2]\tr( \calA_a\sigma_A)
\tr(\calB_b\sigma_B)$, which agrees with the second term of
(\ref{hirsz}). Adding up all the terms we eventually arrive at
(\ref{hirsz}).

One might worry that the resulting state (\ref{stan_z_POVM}) will
never be entangled so its locality will always be trivial. To show
that this is not the case it suffices to apply to above method to
the state\footnote{The corresponding local model for this states
is provided in Section \ref{werner-model}}
$\varrho_0=(1/2)\proj{\psi_-}+(1/4)\proj{0}\otimes \jedynka_2$.
Moreover, this state has an interesting property:
despite having a local model, it displays hidden
non-locality when subject to sequences of measurements
\cite{Hirsch}. More details about the concept of hidden
nonlocality are given in Section \ref{other} below.

Let us finally comment that, in principle, the construction described above can also
be applied to the multiparty scenario. We discuss this possibility in Section
\ref{multiparty-dyskusja}.

\section{Multipartite quantum states}\label{multiparty}
Let us now move to the multipartite scenario. In
this case we will mostly be interested in genuinely multipartite
entangled states for two main reasons. First of all, it is trivial
to construct an $N$-party entangled state which is not GME but has
a local model: it is enough to take the tensor product of a
bipartite state with a local model for two of the parties and a
product state for the remaining $N-2$ parties. Second, any
entangled state which is not GME has a notion of locality, as it
can always be decomposed into a probabilistic mixture of states
that are separable with respect to some bipartition (see Eq.
\ref{biseparowalny}). This, in turn, allows one to directly
construct a hybrid local model for such state combining different
local models for these bipartitions.

Unfortunately, to our knowledge, the literature on the subject is
very limited and boils down to a single local model for a
three-qubit GME state, which we discuss below, and its recent
extension, which we mention in Section \ref{multiparty-dyskusja}.
The question about existence of local models in the general
multipartite setup remains open and at this moment it is far from
clear whether there exist local $N$-partite GME states for $N\ge
4$.

\subsection{Local model for projective measurements on GME tripartite states}\label{trzy-kubity-model}
In what follows we will recall the result of Ref. \cite{TothAcin}
showing that there exist three-qubit GME states with a
local model for projective (two-outcome)
measurements. Here, we  present this result in a slightly different manner
than in the original work \cite{TothAcin}.

Let us consider three parties $A$, $B$, and $C$
and the following simple extension of Werner's
model in which parties $A$ and $B$ behave as in the bipartite
case, and the additional party, Charlie (C), applies the
same strategy as Bob.

Our goal is to show that this local model simulates the outcome probabilities of
projective measurements performed on some GME
tripartite state. To simplify the problem we assume that all the parties
perform two-outcome measurements with measurement operators acting on
$\mathbbm{C}^2$, in other words we aim at obtaining a three-qubit state.
Accordingly, the shared randomness is represented by normalized vectors
$\ket{\lambda}\in\mathbbm{C}^2$ sampled with the probability distribution
$\omega_2(\lambda)$.

In more precise terms, we ask if there exists a three-qubit state
$\rho$ such that the probability
$p_Q(a,b,c|\calA,\calB,\calC)=\Tr[(P_a\ot Q_b\ot R_c)\rho]$ can be
written as
\begin{eqnarray}\label{ProbABC}
 \hspace{-2cm}p_Q(a,b,c|\calA,\calB,\calC)=p_L(a,b,c|\calA,\calB,
\calC)\equiv\int
{\mathrm{d}\lambda}\,\omega_2(\lambda)p(a|\calA,\lambda)p(b|\calB,
\lambda)p(c|\calC,\lambda),\non
\end{eqnarray}
where the response functions of Alice, Bob, and Charlie are given
by (\ref{ProbAA}), (\ref{ProbB}) and
\begin{equation}
p(c|\calC,\lambda)=\langle \lambda|R_c|\lambda\rangle,
\end{equation}
respectively, with local measurements operators $P_a$, $Q_b$, and
$R_c$ $(a,b,c=0,1)$ acting on $\mathbbm{C}^2$.

To verify that this is the case, let us first compute the above
integral. Then we will argue that there exists a quantum state
giving such predictions. It will be particularly useful to exploit
the Bloch representation of one-qubit quantum states (see Section
\ref{Sec:Measurement}). Let then $\boldsymbol{\mathrm{p}}_a$,
$\boldsymbol{\mathrm{q}}_b$ and $\boldsymbol{\mathrm{r}}_c$ denote
the Bloch vectors representing, respectively, $P_a$, $Q_b$, and
$R_c$ on the Bloch sphere\footnote{Recall from Section
\ref{preliminaries} that we need a single vector to characterize a
dichotomic measurement. For notational convenience, however, we do
not exploit this fact here.} and let $\boldsymbol{\lambda}$
represent $\proj{\lambda}$. Exploiting then the fact that
$\langle\lambda|P_a|\lambda\rangle=(1/2)(1+\boldsymbol{\mathrm{p}}
_a\cdot\boldsymbol { \lambda})$ etc., Eq. (\ref{ProbABC}) can be
rewritten as
\begin{eqnarray}
p(a,b,c|\calA,\calB,\calC)=\frac{1}{16\pi}\int_{\boldsymbol{\mathrm{p}}
_a\cdot\boldsymbol{\lambda}<0
}
\mathrm{d}\boldsymbol{\lambda}\,(1+\boldsymbol{\mathrm{q}}_b\cdot\boldsymbol{
\lambda }
)(1+\boldsymbol{\mathrm{r}}_c\cdot\boldsymbol{\lambda}),
\end{eqnarray}
where the fact that the integration is taken over the half-sphere
given by $\boldsymbol{\mathrm{p}}_a\cdot\boldsymbol{\lambda}<0$
stems from the condition in (\ref{ProbAA}). After computing all
integrals (see \ref{appendix} for details), the probabilities can be
finally expressed in terms of the Bloch vectors
$\boldsymbol{\mathrm{p}}_a$, $\boldsymbol{\mathrm{q}}_b$, and
$\boldsymbol{\mathrm{r}}_c$ as
\begin{equation}\label{probsAT}
 p(a,b,c,|\calA,\calB,\calC)=\frac{1}{8}-\frac{1}{16}\left(\boldsymbol{\mathrm{p
}}_a\cdot\boldsymbol{\mathrm{q}}_b+\boldsymbol{\mathrm{p}}_a\cdot\boldsymbol{
\mathrm{r}} _c\right)
+\frac{1}{24}\boldsymbol{\mathrm{q}}_b\cdot\boldsymbol{\mathrm{r}}_c.
\end{equation}
Now, it is fairly easy to construct a three-partite state
realizing these probabilities. First, we notice that a scalar
product of two normalized vectors
$\boldsymbol{\mathrm{x}},\boldsymbol{\mathrm{y}}\in\mathbbm{R}^3$
can be expressed in the ``quantum-mechanical'' form as
\begin{equation}
\boldsymbol{\mathrm{x}}\cdot\boldsymbol{\mathrm{y}}=\sum_{i=1}^3\Tr(\calP
\sigma_i\ot \calQ\sigma_i)=\Tr\left[(\mathcal{P}\ot
\mathcal{Q})\sum_{i=1}^{3}\sigma_i\ot\sigma_i\right],
\end{equation}
where $\mathcal{P}$ and $\mathcal{Q}$ are projectors
corresponding, respectively, to $\boldsymbol{\mathrm{x}}$ and
$\boldsymbol{\mathrm{y}}$ in the Bloch representation.
Applying the above rule to each scalar product in (\ref{probsAT}),
one sees that a state that leads to probabilities (\ref{probsAT})
takes the form
\begin{eqnarray}\label{stan-gme}
\hspace{-2.2cm}
\rho_{ABC}=\frac{\mathbbm{1}_8}{8}-\frac{1}{16}\sum_{i=1}^{3}
\left(\sigma_i\ot\sigma_i\ot\mathbbm{1}_2+\sigma_i\ot
\mathbbm{1}_2\ot\sigma_i\right)
+\frac{1}{24}\sum_{i=1}^3\mathbbm{1} _2\ot\sigma_i\ot\sigma_i.
\end{eqnarray}
It remains to show that it is genuinely multipartite entangled.
The state is manifestly invariant under $U\ot U\ot U$,
$U\in\mathrm{SU}(2)$, since $\sum_i \sigma_i\ot \sigma_i$ is $U\ot
U$ invariant. This allows us to apply the necessary and sufficient
criteria developed in \cite{Eggeling} which confirm that
$\rho_{ABC}$ does indeed contain genuine multipartite
entanglement. Interestingly, the state is a symmetric extension of
the two-qubit Werner state.

\begin{rem} A slight modification of the model allows us
to extend the states (\ref{stan-gme}). Precisely, let us assume that instead of
$\boldsymbol{\lambda}$, the party $A$ uses in their response function a modified
Bloch vector given
by $\boldsymbol{\lambda}'=[a_1\lambda_1,a_2\lambda_2,a_3\lambda_3]$
with some fixed parameters $-1\leq a_i\leq 1$ with $i=1,2,3$
(notice that $\|\boldsymbol{\lambda}'\|\leq 1$).

By repeating the above calculations, one then finds that
\begin{equation}\label{probsAT2}
 p(a,b,c,|\calA,\calB,\calC)=\frac{1}{8}-\frac{1}{16}\left(\boldsymbol{\mathrm{p
}}'_a\cdot\boldsymbol{\mathrm{q}}_b+\boldsymbol{\mathrm{p}}'_a\cdot\boldsymbol{
\mathrm{r}} _c\right)
+\frac{1}{24}\boldsymbol{\mathrm{q}}_b\cdot\boldsymbol{\mathrm{r}}_c
\end{equation}
with $\boldsymbol{\mathrm{p}}_a'$ being $\boldsymbol{\mathrm{p}}_a$ with
coordinates multiplied by $a_i$. These probabilities correspond to the
following class of states
\begin{eqnarray}\label{class}
\hspace{-2cm}\rho_{ABC}'=\frac{\mathbbm{1}_8}{8}-\frac{1}{16}\sum_{i=1}^{3}a_i
\left(\sigma_i\ot\sigma_i\ot\mathbbm{1}_2+\sigma_i\ot
\mathbbm{1}_2\ot\sigma_i\right)
+\frac{1}{24}\sum_{i=1}^3\mathbbm{1} _2\ot\sigma_i\ot\sigma_i.
\end{eqnarray}
Although these states are in general no longer $U\ot U\ot
U$-invariant, the criteria from Ref. \cite{Eggeling} can still be
applied and it follows that whenever $2<a_1+a_2+a_3\leq 3$ holds
$\rho_{ABC}'$ are GME states. It is worth mentioning that for
$c:=a_1=a_2=a_3$, (\ref{class}) reproduces the one-parameter class
of states considered in Ref. \cite{TothAcin} which, as it follows
from the above, is GME if $1\geq c>2/3$. This, in turn, provides an
improvement over the range $1\geq c>(\sqrt{13}-1)/3$, found
originally in \cite{TothAcin}.

Clearly, along the same way one can also modify the Bloch vector
$\boldsymbol{\lambda}$ at Bob's and Charlie's sites. This would
give us some additional parameters in the state (\ref{class}).
\end{rem}

\begin{rem} Let us also notice that the above model
works also if generalized measurements with any number of outcomes
are allowed at Bob's and Charlie's sites. The argument in favor of
this fact is the same as in the case of the Werner states (see
Remark \ref{remark-werner} in Section \ref{werner-model}). This
makes it, in particular, possible to construct more examples of
three-partite states (not necessarily GME) following the reasoning
outlined in Remark \ref{remark_channels} in Section \ref{baret}.
That is, any state of the form $\sigma_{ABC}=\calI_A
\otimes\Lambda_B\ot\Lambda_C(\rho_{ABC})$ with $\Lambda_X$
$(X=A,B)$ being some quantum channels has also a local model for
projective measurements at Alice's site and generalized
measurements at Bob's and Charlie's sites.
\end{rem}

\subsection{Generalizations and
discussion}\label{multiparty-dyskusja}

Let us now discuss how one could apply the results described in
previous sections for two parties to the multiparty scenario.

As already mentioned in Section \ref{pm-to-povm}, the method
described there for the construction of states with a local model
for general measurements from states with a local model for
two-outome projective measuremenrs can be automatically
applied to a multiparty scenario. More
precisely, assume an $N$-partite state $\rho_{\boldsymbol{A}}$
acting on $(\mathbbm{C}^d)^{\ot N}$ has a local model for
projective measurements on each of the sites $A^{(i)}$. Then, the
following state has a local model for arbitrary measurements
\beqn \sigma_{\boldsymbol{A}}=\frac{1}{d^N}\Big\{
\rho_{\boldsymbol{A}}
&+&(d-1)\sum_{i}\rho_{\boldsymbol{A}\setminus
A^{(i)}}\otimes\sigma_{A^{(i)}}\non &+&(d-1)^2\sum_{i<j}
\rho_{\boldsymbol{A}\setminus A^{(i)}A^{(j)}}\otimes
\sigma_{A^{(i)}}\otimes \sigma_{A^{(j)}}\non &+&\cdots\non
&+&(d-1)^N\bigotimes_{i=1}^N \sigma_{A^{(i)}} \Big\},
\eeqn
where $\rho_{\boldsymbol{A}\setminus A^{(i)}}\equiv
\tr_{A^{(i)}}\rho_{\boldsymbol{A}}$ etc. are the reduced states of
$\rho_{\boldsymbol{A}}$, and $\sigma_{A^{(i)}}$ are arbitrary
states acting on $\mathbbm{C}^d$.

The local model for $\sigma_{\boldsymbol{A}}$ is the same as in
Section \ref{pm-to-povm}. Assuming that measurement operators for
party $A^{(i)}$ are $\calA^{(i)}_k=x^{(i)}_k P^{(i)}_k$,
$k=0,1,\cdots,d-1$, it goes as follows. With
probability each party $x^{(i)}_k/d$ chooses
$P^{(i)}_k$ and simulates
dichotomic measurement with elements $P^{(i)}_k$ and
$\jedynka-P^{(i)}_k$ and the corresponding outcomes $+1$ and $-1$.
If the result $+1$ is obtained, the party outputs $x^{(i)}_k$;
otherwise, arbitrary $x^{(i)}_l$ is outputted with probability
$\tr (\sigma_{A^{(i)}}\calA^{(i)}_l)$. Direct calculation confirms
validity of the model for $\sigma_{\boldsymbol{A}}$.

At the moment, the question of whether this construction may lead
to GME states with a local model remains open. Interestingly,
however, a slightly modified approach allows one \cite{Tulio} to
obtain a tripartite qutrit-qubit-qubit state with a local model
for generalized measurements starting from the qubit state given
by T\'{o}th and Ac\'{i}n \cite{TothAcin} (see Section
\ref{trzy-kubity-model}).

A priori, another interesting question is whether it is possible
to apply the machinery based on the \grod constant (see Section
\ref{groth}) to the multipartite scenario. Unfortunately the
corresponding universal constant does not exist even for three
parties. In particular, it is known that in the multiuser case it
is not possible to bound the violation of general correlation Bell
inequalities for dichotomic observables \cite{PerezGarcia}.

Finally, one is tempted to generalize the approach
used in Section \ref{trzy-kubity-model} for the construction of a
GME tripartite state with a local model to a larger number of
parties by appending more parties that behave as in Werner's
original model. It turns out, however, that even for $N=4$ the
resulting state fails to be $U^{\otimes 4}$ invariant so the
straightforward generalization of the tripartite case is not
possible \cite{TothAcin}. Moreover, it has also been proven in
\cite{TothAcin} that in fact there is no four-partite $U^{\otimes
4}$ invariant state for which the model with response functions
for $A$, $B$, and $C$ taken as in Section \ref{trzy-kubity-model}
and for $D$ the same as $B$ and $C$ works. Finally, the tripartite
extension of Werner's model for systems of dimension larger than
two does not work either \cite{TothAcin}.

\section{Other locality scenarios}
\label{other}

In this work, we have reviewed the state of the art concerning
local models for entangled states in the original and more
standard Bell scenario where the parties perform local
measurements on a single copy of the entangled state. It is
however possible to consider other scenarios in which the parties
have access to a larger set of operations and where different
notions of locality appear. In these alternative scenarios, states
that have a local model in the standard scenario might display
nonlocal correlations. Without going into details, we highlight in
what follows some of the known results in these alternative
scenarios:

\begin{itemize}
\item \textbf{Copies of the state:} the first possibility consists of a scenario
in which the parties have access to several copies of a state and
can perform joint measurements on the copies of their subsystems.
In this scenario, it has been shown that nonlocality can be
activated: there exist states $\rho$ that have a local model,
nevertheless, $\rho^{\otimes k}$ is nonlocal for sufficiently
large $k$.  Leaving aside some previous partial results for some
specific inequalities (e.g., \cite{peres-aktywacja,masanes-asymp};
see also \cite{liang,navascues} for other activation scenarios),
the proof of this result is due to Palazuelos \cite{palazuelos}.
His result was later
 generalized in \cite{dani}, where it
was shown that all bipartite entangled states that are useful for
teleportation are nonlocal when considering an arbitrary large
number of copies.
\item \textbf{Network approach:} similar to the previous case but a more general
scenario was
introduced in \cite{network} and consists of a setup in which $k$
copies of a state, say bipartite, are distributed among $N$
parties. Now, one has to test whether the resulting $N$-partite
state, made of $k$ copies of a bipartite state, is nonlocal. In
this scenario, activation effects are also possible in the sense
that there exist networks of local states that display nonlocal
correlations. Moreover, all states that have one-way distillable
entanglement are nonlocal in this scenario \cite{network}.

\item \textbf{LOCC pre-processing:} given a state, it is possible to perform
some
LOCC (local operations and classical communication) pre-processing
before running the Bell test. While classical communication seems
at odds with the concept of nonlocality, this is not the case if
it takes place before the measurements to be performed have been
decided. That is, under this sequential arrangement classical
communication does not create any nonlocality. Bell tests
including LOCC preprocessing were introduced by Popescu in
\cite{popescu} (see also \cite{hiddengis}), who proved that some
local Werner states for $d\ge 5$ become nonlocal in this scenario.
Popescu coined the term \textit{hidden nonlocality} to describe
this phenomenon. While Popescu proved the existence of hidden
nonlocality for states that have a local model for projective
measurements, his result has been generalized to general
measurements in \cite{Hirsch}.
\item \textbf{Copies together with LOCC pre-processing:} finally, one can also
consider the combination
of all the previous possibilities. The resulting scenario is
basically the same as studied for entanglement transformations,
and, in particular entanglement, distillability. Here the goal is
to distill with LOCC from copies of a given state a new state that
violates a Bell inequality. It is clear that in this more general
scenario, all entanglement distillable states are nonlocal. The
natural question is whether nondistillable states, such as states
with positive partial transposition (PPT) are also nonlocal. This
question, know as the Peres conjecture \cite{peres}, was first
proven to be false by D\"ur in the multipartite case \cite{dur} (see also
Refs. \cite{Kaszlikowski,Sen,RAPH,BV}). The
more challenging bipartite case remained open until recently, when
V\'ertesi and Brunner showed the existence of a PPT state violating
a Bell inequality \cite{pereswrong}.
\end{itemize}

As this brief overview shows, there exist several operationally
meaningful notions of nonlocality. What is remarkable is that in
all these scenarios it is still open whether it is possible to
derive an equivalence between entanglement and nonlocality (while
this equivalence is known to fail in the standard scenario).

\section{Conclusions and outlook}

We finish our review with a summary of the main results we have
considered and a short discussion about open questions.

As to the first part, we feel that the best way to provide such a
summary is to recall in a systematized way the constructions
covered in the main body of the paper. We thus have:
\begin{itemize}

\item Werner's model (Section \ref{werner-model}): proof of existence of a
local model for projective measurements (PM) on bipartite
entangled states. The model was constructed for the Werner states
(\ref{WernerState}) in the region
\begin{equation*}
 p\le \frac{d-1}{d},
\end{equation*}

\item Barrett's model (Section \ref{baret}): proof of existence of a local
model for generalized measurements (POVM) on entangled bipartite
states. The model was constructed for the Werner states in the
region
\begin{equation*}\label{korniszony}
 p\leq
 \frac{3d-1}{d(d+1)}\left(\frac{d-1}{d}\right)^{d-1},\nonumber
\end{equation*}
\item the Grothendieck constant approach (Section
\ref{groth}): use of the constant to study the robustness of nonlocal
correlations of entangled bipartite states. In particular:
    \begin{itemize}
        \item there exists a model for PM on a two qubit Werner
        state $\rho_W(2,p)$ if and only if $p \le K_G(3)$, where $K_G(3)$ is the
        Grothendieck constant of order $3$; in particular,
        $\rho_W(2,p)$ is local at least up to $0.6595$,
        \item $\rho_W(2,p)$ is local for planar (at least at one site) PM up to the CHSH threshold
        $p=1/\sqrt{2}$,
        \item for every noisy state
        \beq\label{zaszumiony}
        \sigma=
        p\varrho+(1-p)\frac{\mathbbm{1}_{d^2}}{d^2},
        \eeq
         there is a local model for joint correlations of traceless two-outcome observables for $p<1/K_G(2d^2)$
        for any $\varrho$,
        \item for $p\ge 1/K_G(2\lfloor \log d\rfloor+1)$ there exists a
        state of the form (\ref{zaszumiony}) whose joint
        correlations cannot be reproduced,
        \item (stems from the previous two facts) in the limit
        $d\to\infty$ a state (\ref{zaszumiony}) with arbitrary $\varrho$ and $p\le
        1/K_G$ ($K_G$ -- the \grod constant) is local and this number gives the ultimate limit,
        \item there is a local model simulating full probability
        distribution for traceless observables for the
        isotropic state whenever $p/K_G(d^2-1)$ (see also below),
    \end{itemize}
    \item Almeida {\it et al.}'s construction (Section
    \ref{almeida}): local model for the isotropic states (\ref{isotropic}) and general results for the
    nonlocality of noisy entangled states (\ref{zaszumiony}). One has:
    \begin{itemize}
        \item a local model for PM on the isotropic states (\ref{isotropic}) for

        \begin{equation*}\nonumber
        p\le p_{\mathrm{PM}}^{\mathrm{iso},c}\equiv \frac{\sum_{k=2}^d
        \frac{1}{k}}{d-1},
        \end{equation*}
        \item a model for POVM on the isotropic states (\ref{isotropic}) for
        %$p\le p_{\mathrm{POVM}}^{\mathrm{iso,c}}$ with
%
         \begin{equation*}
      p\le p_{\mathrm{POVM}}^{\mathrm{iso,c}}\equiv \frac{3d-1}{d(d+1)}\left(\frac{d-1}{d}\right)^{d-1},\nonumber
      \end{equation*}

        \item a local model for PM on arbitrary states of the form
        (\ref{zaszumiony}) for
        \begin{equation*}
        p\le\frac{p^{\mathrm{iso},c}_{\mathrm{PM}}}{(1-p^{\mathrm{iso},c}_{\mathrm{PM}})(d-1)+1},\nonumber
        \end{equation*}
        \item a local model for POVM on arbitrary states of the form (\ref{zaszumiony}) for
         \begin{equation*}
        p\le\frac{p_{\mathrm{POVM}}^{\mathrm{iso,c}}}{(1-p_{\mathrm{POVM}}^{\mathrm{iso,c}})(d-1)+1},\nonumber
        \end{equation*}
    \end{itemize}
    \item Hirsch \etal's construction (Section \ref{pm-to-povm}): novel
examples of local states. The work
    provides a systematic method to derive states local for POVMs from states local
    for PMs of two outcomes. Given a state
$\varrho_0$ local
    for dichotomic PM one builds the state $\varrho=(1/d^2)[
\varrho_0+(d-1)
(\varrho_A\otimes\sigma_B+\sigma_A\otimes\varrho_B)+
(d-1)^2\varrho_A\otimes\sigma_B ]$ with $\varrho_{A,B}$ being
reductions of $\varrho_0$ and $\sigma_{A,B}$ arbitrary, which is
local for POVM,
    \item T\'oth-Ac\'in model (Section \ref{trzy-kubity-model}): proof of
existence of genuinely entangled three-qubit states with a local
model for PMs. The model is based on a symmetric extension of the
two-qubit Werner model.
   Building on this result, the existence of three-partite GME
states local under general measurements has been proven recently \cite{Tulio}.
\end{itemize}
It is worth mentioning that all the previous construction are, in
a way or another, based on Werner's original model. In this sense,
it would be interesting to have a model that does not use at any
point some of the ingredients in Werner's model.

Finally, let us conclude with a brief discussion of the main open questions
in the relation between entanglement and nonlocality. In the case of two parties,
we can think of the following three main questions:
\begin{enumerate}
\item It is known that in the standard Bell scenario, entanglement and nonlocality are
inequivalent. Is it possible to identify a more general scenario
in which these two quantum phenomena become equivalent?
\item For many of the family of states considered above, there always appear a gap
between the region where a local model exists for projective
measurements and the one for general measurements. It is an open
question whether this gap is an artifact of the construction or it
actually exists. That is, do general measurements offer any
advantage to detect the nonlocality of quantum states?
\item An almost unexplored question is to study local models from a quantitative
point of view. The existence of a local model for a quantum state operationally means
that the correlations of the state can be reproduced using shared randomness. How
much shared randomness is needed? How does it scale with the entanglement in
the state?
\end{enumerate}

Moving to the multipartite case, basically all questions remain
open. To our knowledge, there is not a single example of a local
model for an entangled state of more than three parties. But,
perhaps the most interesting questions concern whether gaps exist
for any number of parties. In this direction, we can identify the
following questions:
\begin{enumerate}
\item The first question concerns whether
for any number of parties it is possible to find GME states that
have a fully local model. We do not have any strong intuition
about this. In principle, based on all the existing results, one
may expect a gap for any number of parties. However, when going to
many parties, it could be that GME are so entangled that a fully
local model cannot reproduce their statistics. If this was the
case, it would be interesting to estimate the minimal number of
parties such that local models do not exist. Thus, the question
is: is there a finite number of parties such that all GME violate
a standard Bell inequality?
\item Perhaps a more fair comparison between entanglement and
nonlocality in the multipartite scenario consists of comparing GME
with genuinely multipartite nonlocality. Here, we are interested
in understanding whether there exist for any number of parties GME
states that can be described by a local model in which some of the
parties join, as first considered by Svetlichny \cite{Svetlichny}
(see also \cite{operational,operational2}). Thus, the question is:
is there a finite number of parties such that all GME violate a
Svetlichny-Bell inequality? While finishing writing this
manuscript, we have proven that the answer to this question is
negative. That is, for any number of parties, we have found GME
states that can be described by local models in which subset of
the parties join \cite{My}.
\end{enumerate}

After all these discussions, the main message of this review
becomes clear: 50 years after Bell's seminal paper \cite{Bell} we
are still very far from understanding the relation between
entanglement and nonlocality.

\ack

 This work is supported by the EU project SIQS, ERC CoG QITBOX, the Spanish
Chist-Era DIQIP project. This publication was made possible
through the support of a grant from the John Templeton
Foundation. R. A. also acknowledges the Spanish MINECO for the support through
the Juan de la Cierva program.

\appendix
\section{}
\label{appendix}
%{ Computation of the integrals}

Here we compute analytically the integrals that have been
introduced in the main text.

\paragraph{The integral $J[u_1]$ in Eq. (\ref{jot-u-jeden}).}
Here, following Mermin \cite{Mermin}, we will show that
$J[u_1]=N/d^3$. Recall that its explicit form reads
\begin{equation}
\int_0^1\mathrm{d}u_1\int_{u_1}^1\mathrm{d}u_2\ldots
\int_{u_1}^1\mathrm{d}u_d\, u_1 \,\delta(u_1+\ldots +u_d-1).
\end{equation}
Without changing the value of this integral but to simplify its
computation we can extend the range of all the variables to
infinity, i.e.,
\begin{equation}\label{appA:1}
\int_0^{\infty}\mathrm{d}u_1\int_{u_1}^{\infty}
\mathrm{d}u_2\ldots \int_{u_1}^{\infty}\mathrm{d}u_d\, u_1
\,\delta(u_1+\ldots +u_d-1).
\end{equation}
By changing the variables $u_i=v_i+u_1$ for $i=2,\ldots,$ and then
$u_1=v_1/d$, it can further be rewritten as
\begin{equation}
\frac{1}{d^2}\int_0^{\infty}\mathrm{d}v_1
\ldots \int_{0}^{\infty}\mathrm{d}v_d\, v_1 \,\delta(v_1+\ldots
+v_d-1).
\end{equation}
One finally notices that the value of the integral is the same if
$v_1$ is replaced by any $v_i$, and consequently
\begin{eqnarray}
&&\frac{1}{d^2}\int_0^{\infty}\mathrm{d}v_1
%
%\int_{0}^{\infty}\mathrm{d}v_2
%
\ldots \int_{0}^{\infty}\mathrm{d}v_d\, v_1 \,\delta(v_1+\ldots
+v_d-1)\nonumber\\
&&=\frac{1}{d^3}\int_0^{\infty}\mathrm{d}v_1
%
%\int_{0}^{\infty}\mathrm{d}v_2
%
\ldots \int_{0}^{\infty}\mathrm{d}v_d\, (v_1+\ldots+v_d)
\,\delta(v_1+\ldots
+v_d-1)\nonumber\\
&&=\frac{1}{d^3}N,
\end{eqnarray}
which is what we wanted to prove. Recall that $N$ is defined by
Eq. (\ref{N}).

\paragraph{The integral $\widetilde{J}[u_1]$ in Eq. (\ref{jot-u-jeden-tylda}).}
Now, repeating the same tricks
as above, we compute
\begin{equation}
\widetilde{J}[u_1]=\int_{1/d}^1\mathrm{d}u_1\int_{0}^1\mathrm{d}u_2\ldots
\int_{0}^1\mathrm{d}u_d \,u_1\,\delta(u_1+\ldots +u_d-1).
\end{equation}
As before, we can exploit the fact that the argument of
integration contains Dirac delta and extend upper bounds of all
integrals to infinity. Then, by a change of the variable $u_1\to
u_1+1/d$, one gets
\begin{eqnarray}
\widetilde{J}[u_1]&=&\int_{0}^{\infty}\mathrm{d}u_1\ldots
\int_{0}^{\infty}\mathrm{d}u_d \,u_1\,\delta(u_1+\ldots
+u_d-\case{d-1}{d})\nonumber\\
&&+ \frac{1}{d}\int_{0}^{\infty}\mathrm{d}u_1\ldots
\int_{0}^{\infty}\mathrm{d}u_d\, \delta(u_1+\ldots
+u_d-\case{d-1}{d})
\end{eqnarray}
Now, we change all variables $u_i\to u_i(d-1)/d$ with
$i=1,\ldots,d$, and then use the property
$\delta(ax)=(1/|a|)\delta(x)$, which gives
\begin{eqnarray}
\widetilde{J}[u_1]&=&\left(\frac{d-1}{d}\right)^{d}\int_{0}^{\infty}\mathrm{d}
u_1\ldots\int_{0}^{\infty}\mathrm{d}u_d \,u_1\,\delta(u_1+\ldots
+u_d-1)\nonumber\\
&&+
\frac{1}{d}\left(\frac{d-1}{d}\right)^{d-1}\int_{0}^{\infty}\mathrm{d}u_1\ldots
\int_{0}^{\infty}\mathrm{d}u_d\, \delta(u_1+\ldots
+u_d-1).\nonumber\\
\end{eqnarray}
To complete the proof, one notices that the first integral has
been already computed and amounts to $N/d$, while the second one
is simply $N$. As a result,
\begin{equation}
\widetilde{J}[u_1]=\frac{N}{d}\frac{2d-1}{d^2}\left(\frac{d-1}{d}\right)^{
d-1 }.
\end{equation}

\paragraph{The integral $\widetilde{J}[u_1^2]$ in Eq.
(\ref{jot-u-jeden-kwadrat-tylda}).} Following the same steps as
before, it is fairly easy to see that %the integral
\begin{equation}
\widetilde{J}[u_1^2]=\int_{1/d}^1\mathrm{d}u_1\int_{0}^1\mathrm{d}u_2\ldots
\int_{0}^1\mathrm{d}u_d \,u_1^2\,\delta(u_1+\ldots +u_d-1).
\end{equation}
can be expressed as
\begin{eqnarray}
 \widetilde{J}[u_1^2]&=&\frac{N}{d^2}\left(\frac{d-1}{d}\right)^{d-1}+\frac{2N}{
d^2
}\left(\frac{d-1}{d}\right)^{d}+\left(\frac{d-1}{d}\right)^{d+1}\nonumber\\
&&\times\int_{0}^{\infty}\mathrm{d}u_1\ldots
\int_{0}^{\infty}\mathrm{d}u_d \,u_1^2\,\delta(u_1+\ldots +u_d-1).
\end{eqnarray}
The last integral can be computed directly. From the fact that it
contains the Dirac delta, one obtains
\begin{eqnarray}
 \int_{0}^{\infty}\mathrm{d}u_1\ldots
\int_{0}^{\infty}\mathrm{d}u_d \,u_1^2\,\delta(u_1+\ldots +u_d-1)\nonumber\\
=\int_{0}^{\infty}\mathrm{d}u_1u_1^2\int_0^{1-u_1}\mathrm{d}u_2\ldots\int_0^{
1-u_1-\ldots-u_{d-2}}\mathrm{d}u_{d-1}.
\end{eqnarray}
A straightforward integration over the variables
$u_2,\ldots,u_{d-1}$ gives
\begin{eqnarray}
 \int_{0}^{\infty}\mathrm{d}u_1\ldots
\int_{0}^{\infty}\mathrm{d}u_d \,u_1^2\,\delta(u_1+\ldots +u_d-1)\nonumber\\
=\frac{1}{(d-2)!}\int_{0}^1\mathrm{d}u_1u_1^2(1-u_1)^{d-2},
\end{eqnarray}
which further amounts to
%
%The last integration can be performed by substituting $v=1-u_1$, leading to
%
\begin{eqnarray}
 \int_{0}^{\infty}\mathrm{d}u_1\ldots
\int_{0}^{\infty}\mathrm{d}u_d \,u_1^2\,\delta(u_1+\ldots +u_d-1)
=\frac{2N}{d(d+1)},
\end{eqnarray}
and consequently,
\begin{eqnarray}
\widetilde{J}[u_1^2]&=&\frac{N}{d}\left[\frac{1}{d}+\frac{2}{d}\frac{d-1}{d}
+\frac{2}{d+1}\left(\frac{d-1}{d}\right)^2\right]\left(\frac{d-1}{d}\right)^{d-1
}\nonumber\\
&=&\frac{N}{d}\frac{5d-3}{d+1}\left(\frac{d-1}{d}\right)^{d-1}.
\end{eqnarray}

\paragraph{The integral $\mathcal{J}[u_1]$ in Eq. (\ref{jot-cal-u-jeden})} Let
us now compute the following integral
\begin{equation}
\mathcal{J}[u_1]=\int_{0}^1\mathrm{d}u_1
\int_{0}^{u_1}\mathrm{d}u_2\ldots \int_{0}^{u_1}\mathrm{d}u_d
\,u_1\delta(u_1+\ldots +u_d-1).
\end{equation}
For this purpose, we rewrite all integrals over the variables
$u_i$ $(i=2,\ldots,d)$ as
\begin{equation}
\int_{0}^{u_1}\mathrm{d}u_i=\int_{0}^{\infty}\mathrm{d}u_i-\int_{u_1}^{\infty}
\mathrm{d}u_i,
%
%\qquad (i=2,\ldots,d)
\end{equation}
which after some algebra leads us to
\begin{eqnarray}
\mathcal{J}[u_1]&=&\sum_{k=0}^{d-1}
{{d-1}\choose{k}}(-1)^k\int_{0}^{\infty}\mathrm{d}u_1
u_1\int_{u_1}^{\infty}\mathrm{d}u_2\ldots\mathrm{d}u_{k+1}\nonumber\\
&&\times\int_{0}^{\infty}\mathrm{d}u_{k+2}\ldots \mathrm{d}u_{d}\,
\delta(u_1+\ldots +u_d-1),
\end{eqnarray}
where we have exploited the fact that the integrated function is
invariant under any permutation of the variables $u_2,\ldots,u_d$.
Now, by changing the variables $u_i\to u_i-u_1$ with
$i=2,\ldots,k+1$, and then $u_1\to u_1/(1+k)$, each integration in
the summand rewrites as
\begin{eqnarray}
 \int_{0}^{\infty}\mathrm{d}u_1
u_1\int_{u_1}^{\infty}\mathrm{d}u_2\ldots\mathrm{d}u_{k+1}
\int_{0}^{\infty}\mathrm{d}u_{k+2}\ldots
\mathrm{d}u_{d}\delta(u_1+\ldots
+u_d-1)\nonumber\\
 =\int_{0}^{\infty}\mathrm{d}u_1
u_1\int_{0}^{\infty}\mathrm{d}u_2\ldots\mathrm{d}u_{d}
\delta((k+1)u_1+\ldots
+u_d-1)\nonumber\\
=\frac{1}{(1+k)^2}\int_{0}^{\infty}\mathrm{d}u_1
u_1\int_{0}^{\infty}\mathrm{d}u_2\ldots\mathrm{d}u_{d}
\delta(u_1+\ldots +u_d-1)
\end{eqnarray}
The last integral has already been computed and it amounts to
$N/d$, which finally gives
\begin{equation}
 \mathcal{J}[u_1]=\frac{N}{d}\sum_{k=0}^{d-1}
{{d-1}\choose{k}}\frac{(-1)^k}{(1+k)^2}=\frac{N}{d^2}
\sum_{k=1}^{d}\frac{1}{k} .
\end{equation}
%
%\textbf{Show that}
To obtain the last equality we have utilized one of the
representations of the harmonic number $H_d\equiv
\sum_{k=1}^{d}\frac{1}{k}$, namely:
\begin{equation}
H_d=
 \sum_{k=0}^{d-1}
{{d-1}\choose{k+1}}\frac{(-1)^k}{1+k}.
\end{equation}

\paragraph{Integrals over the Bloch sphere.} Here we show that
\begin{equation}\label{BlochInt1}
\displaystyle\int_{\boldsymbol{\mathrm{x}}\cdot\boldsymbol{\lambda}<0}\mathrm{d}
\boldsymbol{\lambda}\,
(\boldsymbol{\mathrm{y}}\cdot\boldsymbol{\lambda})=-\pi
\boldsymbol{\mathrm{x}}\cdot\boldsymbol{\mathrm{y}}
\end{equation}
where $\boldsymbol{\mathrm{x}}$, $\boldsymbol{\mathrm{y}}$, and
$\boldsymbol{\lambda}$ are unit vectors from $\mathbbm{R}^3$.

Let us first note that both integrals are invariant under any
rotation of the Bloch sphere, i.e.,
\begin{equation}
\displaystyle\int_{\boldsymbol{\mathrm{x}}\cdot\boldsymbol{\lambda}<0}\mathrm{d}
\boldsymbol{\lambda}\, (\boldsymbol{\mathrm{y}}\cdot\boldsymbol{\lambda})=
\displaystyle\int_{\boldsymbol{\mathrm{x}}'\cdot\boldsymbol{\lambda}<0}\mathrm{d
}
\boldsymbol{\lambda}\, (\boldsymbol{\mathrm{y}}'\cdot\boldsymbol{\lambda}),
\end{equation}
where $\boldsymbol{\mathrm{x}}'=O\boldsymbol{\mathrm{x}}$ and
$\boldsymbol{\mathrm{y}}'=O\boldsymbol{\mathrm{y}}$ with $O$ being an element of
the $SO(3)$ group. This allows us to choose the reference frame such
that $\boldsymbol{\mathrm{x}}$ is aligned with its $z$-axis, i.e., we
choose such $O$ that $O\boldsymbol{\mathrm{x}}=[0,0,1]$. Now, with respect
to this new reference frame we can parametrize
$\boldsymbol{\lambda}$ as
$\boldsymbol{\lambda}=[\cos\phi\sin\theta,\sin\phi\sin\theta,\cos\theta]$
with $\phi\in[0,2\pi]$ and $\theta\in[0,\pi]$. As a result, the
integral (\ref{BlochInt1}) rewrites as
\begin{equation}
\int_{\pi/2}^{\pi}d\theta\sin\theta\int_{0}^{2\pi}\mathrm{d}\phi\,
(y_1'\cos\phi\sin\theta+y_2'\sin\phi\sin\theta+y_3'\cos\theta),
\end{equation}
where the range of the first integral comes from the condition
$\boldsymbol{\mathrm{x}}\cdot\boldsymbol{\lambda}<0$, which after rotation
simplifies to $\cos\theta<0$. Performing all integrations in the
above, one obtains that
\begin{equation}
\displaystyle\int_{\boldsymbol{\mathrm{x}}\cdot\boldsymbol{\lambda}<0}\mathrm{d}
\boldsymbol{\lambda}\,
(\boldsymbol{\mathrm{y}}\cdot\boldsymbol{\lambda})=-\pi
y_3'=-\pi\boldsymbol{\mathrm{x}}\cdot\boldsymbol{\mathrm{y}}.
\end{equation}

Along exactly the same lines one also finds that
\begin{equation}
\displaystyle\int_{\boldsymbol{\mathrm{x}}\cdot\boldsymbol{\lambda}<0}\mathrm{d}
\boldsymbol{\lambda}\,
(\boldsymbol{\mathrm{y}}\cdot\boldsymbol{\lambda})\,(\boldsymbol{\mathrm{z}}
\cdot\boldsymbol { \lambda})
=\frac{2\pi}{3}\boldsymbol{\mathrm{y}}\cdot\boldsymbol{\mathrm{z}}.
\end{equation}

\section*{References}
\bibliography{cytacje2}
\bibliographystyle{iopart-num}
\end{document}